\documentclass[a4paper,oneside]{article}
\pdfoutput=1
\usepackage{graphicx}
\usepackage[]{indentfirst}
\usepackage[]{amsfonts}
\usepackage[]{amssymb}

\newcommand{\be}{\begin{equation}}
\newcommand{\ee}{\end{equation}}
\newcommand{\bea}{\begin{eqnarray}}
\newcommand{\eea}{\end{eqnarray}}

\begin{document}

\title{Networks of coupled circuits: From a versatile toggle switch to collective coherent behavior}

\author{Darka Labavi\'{c}\footnote{email: d.labavic@jaocbs-university.de} \and Hildegard Meyer-Ortmanns\footnote{email: h.ortmanns@jacobs-university.de}}

\date{\small{School of Engineering and Science, Jacobs University Bremen, P.O.Box 750561, D-28725 Bremen, Germany}}

\maketitle

\begin{abstract} We study the versatile performance of networks of coupled  circuits. Each of these circuits is composed of a positive and a negative feedback loop in a motif that is frequently found in genetic and neural networks. When two of these circuits are coupled with mutual repression, the system can function as a toggle switch. The variety of its states can be controlled by two parameters as we demonstrate by a detailed bifurcation analysis. In the bistable regimes switches between the coexisting attractors  can be induced by noise. When we couple larger sets of these units, we numerically observe collective coherent modes of individual fixed-point and limit-cycle behavior. It is there the monotonic change of a single bifurcation parameter that allows to control the onset and arrest of the synchronized oscillations. This mechanism may play a role in biological applications, in particular in connection with the segmentation clock. While tuning  the bifurcation parameter, also a variety of transient patterns emerges upon approaching the stationary states, in particular a self-organized pacemaker in a completely uniformly equipped ensemble, so that the symmetry breaking happens dynamically.
\end{abstract}

{\bf From the physics' point of view, it is of generic interest how complex the dynamics of individual building blocks should be assumed  in order to reproduce the versatile collective behavior of genetic or neural networks when these blocks are coupled.  As a basic unit of our network of coupled circuits we consider the motif of a self-activating species $A$ that activates its own repressor, a second species $B$. The species stand for the concentrations of genes, proteins or cells. An individual unit behaves like an excitable or oscillatory element, depending on the choice of parameters. So the individual dynamics is more complex than that of phase oscillators or repressilators, but as we later shall see, it allows for a simple control when these units are coupled. When two of these units mutually repress each other, we observe already a proliferation of possible attractors, differing by their amplitude and periods of oscillations, their pattern of synchronized phases, or their fixed point values. Such a system may act like a synthetic toggle switch whose multistable states can be addressed in a controlled way. It is then challenging to explore how a larger network of these units may function. In our numerical simulations of a set of some thousands of these units we observe collective fixed-point behavior for small and large values of a certain bifurcation parameter, and  collective synchronized oscillations for intermediate parameter values. The bifurcation parameter is related to the production rate of species $A$. This behavior allows for a simple control of the onset and arrest of oscillations via tuning  this single parameter through two bifurcations. It is possible without any finetuning or external interference. Such a mechanism seems to be simple enough to be realized in biological applications like the segmentation clock. Moreover we observe an interesting phenomenon of dynamical symmetry breaking: a transient, self-organized pacemaker emerges in our uniformly equipped ensemble of oscillators. It emits target waves for some thousands of time units before the stationary state is reached.  It is then instructive to perform  a detailed bifurcation analysis for two coupled units. Such an analysis illustrates how a varying coupling constant or an increasing system size may lead to complex bifurcation patterns:  pitchfork bifurcations, where fixed points are created or annihilated, fold bifurcations, in which  limit cycles collide and disappear,  or Hopf bifurcations, as they first split and later disappear  under these variations. This explains why it is inherently difficult to analytically anticipate the performance and functioning of larger ensembles in all their versatility, in particular  the option of their simple control.}

\section{Introduction}

Our networks of coupled  circuits are composed of units which are known as a common motif in neural and genetic networks in different realizations. The basic motif is found whenever bistable units are coupled to negative feedback loops. The motif consists of a first species $A$ that activates itself in a positive feedback loop, while it also activates a second species $B$ in a negative feedback loop that acts as its own repressor. Examples are signaling systems like the slime mold Dictyosthelium Discoideum \cite{mold},  the  embryonic division control system \cite{pomerening}, or the MAPK-cascade \cite{mapk}, and the circadian clock \cite{leibler}.

As it was shown in \cite{pablo}, this motif shows three regimes, characterized by different attractors, when a single bifurcation parameter is monotonically increased: excitable behavior, where the system approaches a fixed point (if the perturbation from the fixed point exceeds a certain threshold, the system makes a long excursion in phase space before it returns to the fixed point), limit-cycle behavior, and again excitable behavior. The individual dynamics resembles FitzHugh-Nagumo units \cite{fitzhugh,nagumo}. So it is more complex than bistable systems, phase oscillators or repressilators, and when these units are coupled, it should not come as a surprise that the resulting dynamics can be rather versatile.

Here we consider two cases: a coupled set of minimal size, these are two units with mutually repressive coupling, and a large set of the order of a few thousands of these units, coupled with repression on two-dimensional grids. The first case will illustrate the proliferation of attractors induced by the coupling, and the kind of bifurcations which influence the complex structure of the attractor landscape; the latter one reveals collective coherent behavior that resembles an individual unit and allows for possible applications in genetic or cellular networks.

The system of two units may function as a synthetic device like a toggle switch, for which the system here can toggle between eight different regimes in a two-dimensional parameter space. The parameter space is spanned by the bifurcation parameter $\alpha$, which is related to the production rate of the first species $A$, and the coupling strength $\beta$. The different regimes are characterized by their attractors. The attractors can be fixed-point solutions or synchronized oscillations with different patterns of phase-locked motion. A single regime may have multistable stationary states. In such a situation it is of interest how we can control the return to a given initial state after we perform a closed path in parameter space.

\vskip3pt For the larger system we choose a two-dimensional grid of $50\times 50$ oscillators. As for an individual unit, also the collective behavior of the whole ensemble can be easily tuned via a single (namely the same ) bifurcation parameter to vary from collective fixed-point to synchronized limit-cycle to collective fixed-point behavior. This option exists for  a wide range of parameters for different lattice topologies, boundary conditions and initial conditions. The very patterns of phase-locked motion in the oscillatory regime and the combination of fixed points in the other regimes depend, however,  on the geometry, the choice of free or periodic boundary conditions, and the random or structured initial conditions.

In particular we indicate applications to patterns in the so-called segmentation clock. In relation to the segmentation clock it is an open question which mechanism is responsible in particular for the arrest of genetic oscillations. Our coupled circuits initiate and arrest the oscillations via two bifurcations of the same dynamics, once a single parameter is monotonically changed. Therefore we speculate that the same kind of mechanism may be realized in the segmentation clock, without any need for finetuning  some parameters or relying on external interference.

As a remarkable transient phenomenon we describe the emergence of target waves that are seemingly emitted from a single site on the grid, so that the symmetry  between the uniformly equipped oscillators is dynamically broken. These waves look like a self-organized pacemaker and fade away after several thousand time units, a period that may be sufficiently long to have an impact on biological systems.

The paper is organized as follows. In Sec.~\ref{sec2} we present the model. Sec.~\ref{sec3} is devoted to a detailed bifurcation analysis of the toggle switch and a partial extrapolation of bifurcations to larger system sizes, including some results about the functioning as a toggle switch. In Sec.~\ref{sec4} we summarize the results of numerical simulations for the larger system size and point to possible biological applications. A summary and outlook is given in Sec.~\ref{sec5}. The appendices  contains further details of the bifurcation analysis for the toggle switch.

\section{The model}\label{sec2}
\begin{figure}[h]\label{figure1}
	\begin{center}
		\includegraphics[scale=0.5]{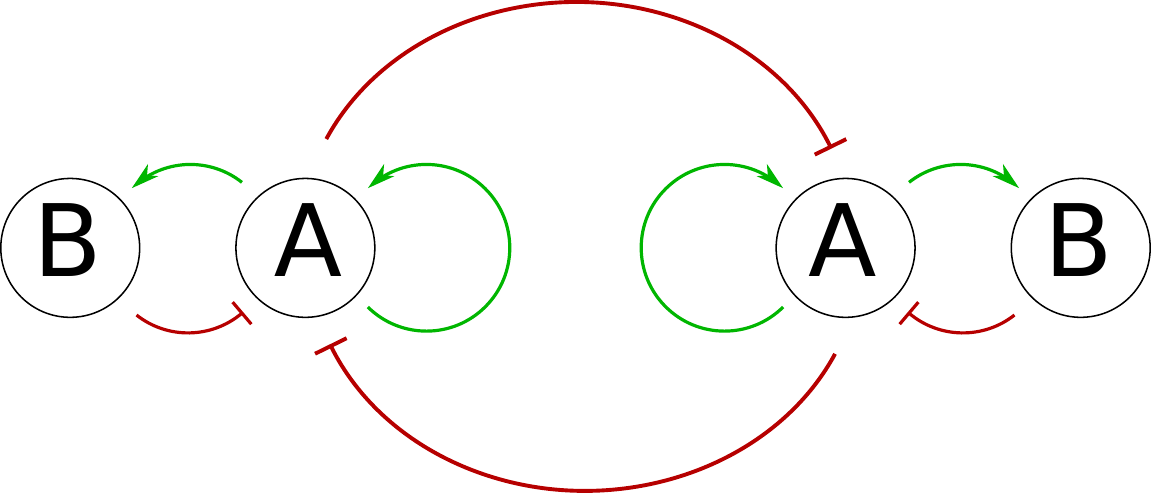}
	\end{center}
	\caption{(Color online) The system of two coupled circuits, repressing each other. Gray (green) arrows represent activation links and dark gray (red) blunts repression links. Species $A$ activates itself as well as species $B$, while species $B$ represses species $A$. The interaction between the two units is realized by repression between two individuals of species $A$.}
\end{figure}

The motif of a single genetic circuit is shown in Figure~\ref{figure1}.

In this model, $A$ and $B$ are two different species. Although from a statistical physics point of view the identity of $A$, $B$ is not important, we shall assume here that $A$, $B$ are two different protein types. This is in line with the original motivation of this model as a coarse-grained description of some genetic circuits. The protein $A$ activates its own production (transcription in the biological language) and also the production of the protein $B$, which in turn represses the production of $A$. In this way, we have a self-activating bistable unit which is coupled to a negative feedback loop.
The simplest, idealized implementation of this unit has been analyzed on a deterministic level, with protein concentrations as the only dynamical variables, and it is known to produce oscillations in a certain range of model parameters \cite{guantes,sandeep}.
In particular, the model studied in Ref.~\cite{sandeep}, which we adopt here, assumes that the protein production rates depend on the concentrations $\phi_A$ and $\phi_B$ of the two protein species as follows:
\begin{eqnarray}
 \frac{d\phi_A}{dt}&=&\frac{\alpha}{1+\phi_B/K}\,\frac{b+\phi_A^2}{1+\phi_A^2}-\phi_A, \label{eq-det11} \\
 \frac{d\phi_B}{dt}&=&\gamma(\phi_A-\phi_B), \label{eq-det12}
\end{eqnarray}
where $\gamma$ is the ratio of the half-life of $A$ to that of $B$. Here we shall focus on the case $\gamma\ll 1$, that is when the protein $B$ has a much longer half-life than $A$ with a slow reaction on changes in $A$, while $A$ has a fast response to changes in $B$.  The parameter $K$ sets the strength of repression of $A$ by $B$. We shall assume $K\ll 1$, so that already a small concentration of $B$ will inhibit the production of $A$. The parameter $b$ determines the basal expression level of $A$. We set this parameter to anything larger than zero but much smaller than one, so that the system cannot be absorbed in the state $\phi_A=\phi_B=0$ and, simultaneously, the production rate of $A$ is small for $\phi_A\approx 0$. The Hill coefficients (powers of $\phi_A$, $\phi_B$ on the r.h.s. of Eqs.~(\ref{eq-det11})-(\ref{eq-det12})) are chosen as in \cite{sandeep,pablo}. We keep them fixed throughout the paper.
The parameter $\alpha$ is the maximal rate of production of $A$ for full activation ($\phi_A^2\gg b$) and no repression ($\phi_B\approx 0$). This will be one of our tunable parameters which we will use to control the behavior of our model, also when these units are coupled. This parameter seems to be also the easiest one to control in real, experimental systems \cite{sandeep}. The different fixed-point and oscillatory regimes of such an individual unit are separated  by subcritical Hopf bifurcations (for a definition see for example Ref.~\cite{strogatz}) with corresponding hysteresis effects \cite{pablo}.

From now on, we do no longer distinguish between the species and their concentrations in the notation and write $A,B$ for the corresponding concentrations. We consider coupled units with only repressing couplings (for simplicity, since the effect of activating couplings can be compensated by an appropriate choice of geometry, see \cite{pablo}) according to
\begin{eqnarray}\label{eq3}
&&\frac{dA_i}{dt}\;=\;\frac{\alpha}{1+(B_i/K)}\;\cdot\;\left(\frac{b+A_i^2}{1+A_i^2}\right)\;-\;A_i\\ \nonumber
&&+\beta\;\sum_{j=1}^{N}R_{ij}\frac{1}{1+(A_j/K)^2}\\
&&\frac{dB_i}{dt}\;=\;\gamma (A_i\;-\;B_i)\;,\qquad i=1,...,N\;.\nonumber
\end{eqnarray}
Here $i$ labels the units.  The parameters $\alpha$, $K$, $b$ and $\gamma$ are chosen uniformly over the grid, and $K$, $b$ and $\gamma$ with the same values as introduced for a single unit, that is $K=0.02$, $b = \gamma=0.01$, so without any finetuning, and kept fixed throughout the paper like the Hill coefficients. We should remark that the bifurcation diagram and the synchronization properties may alter if one of these parameters would change. For a system of two coupled repressilators it was shown \cite{potapov} that the very Hill coefficients and reaction time scales (in our case also implicitly kept fixed) can dramatically influence the type of synchronization, so that it is not the mere network topology of activating or repressing couplings that determines the synchronization. Therefore our conclusions should be understood  to hold for given fixed values of these parameters. The parameter $\alpha$ is also chosen independently of the lattice site, but as a bifurcation parameter it is varied between $1$ and $110$ or $400$, depending on the repressive coupling strength, parameterized by $\beta$, being weak ($\beta=0.1$) or strong ($\beta=10.0$), respectively. In general $\beta$ is chosen with the same value on all bonds and in opposite directions, although we shall briefly discuss the effect of choosing it asymmetrically in the toggle switch in section \ref{secbifurcation}.
Finally, the adjacency matrix elements $R_{ij}$ take values 1 if there is a directed coupling from $i$ to $j, i\not=j$ and zero otherwise. The adjacency matrix will be chosen to represent different lattice topologies as explained later in the corresponding sections.

\section{A Toggle Switch of Two Genetic Circuits}\label{sec3}

\subsection{The Bifurcation Diagram}
\begin{figure}\label{figure2}
	\begin{center}
		\includegraphics[width=8cm]{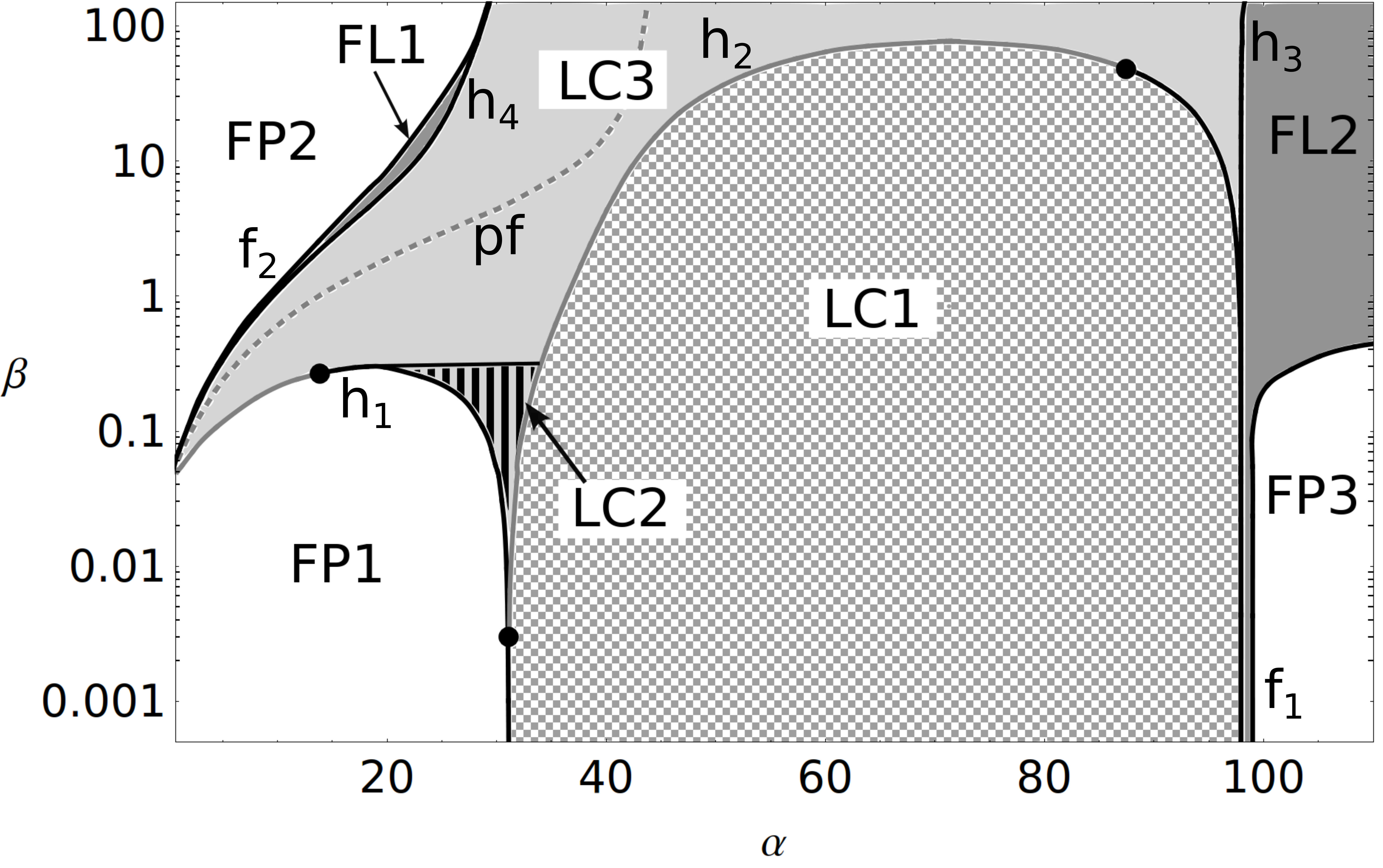}
	\end{center}
	\caption{Bifurcation diagram. Black full lines represent subcritical Hopf bifurcation ($h_i$) or fold bifurcations of limit cycles ($f_i$), gray full lines represent supercritical Hopf bifurcations, and gray dotted lines represent pitchfork bifurcations ($pf$). Black points stand for transition points from subcritical to supercritical Hopf bifurcations. The shaded areas inside the bifurcation branches represent different regimes of attractors: $FPi$ stand for single or coexisting fixed points, $FLi$ for the coexistence of one or two fixed points and a limit cycle, and $LCi$ for one or coexisting limit cycles with indices $i$ labeling the corresponding regimes. Fur further explanations see the text.}
\end{figure}

In the following we encounter fold bifurcations, sub- or supercritical pitchfork bifurcations and sub- or supercritical Hopf bifurcations. In our four-dimensional system a Hopf bifurcation may amount to the lost of stability in all rather than in only two directions. This can lead to the birth  of saddle-limit cycles rather than stable or unstable limit cycles.
Depending of the choice of $\alpha$ and $\beta$,  the system can then be found in eight different regimes, as summarized  in Figure~\ref{figure2} and reviewed in the following.
The diagram is based on data points that are continued to solid or dashed lines, indicating the lines of bifurcation. The attractors in these regimes are either fixed points ($FP$) (single or coexisting ones), limit cycles ($LC$) (also single or coexisting ones), and limit cycles coexisting with fixed points ($FL$).

\subsubsection{Fixed point regimes}
\begin{figure}\label{figure3}
	\begin{center}
		\includegraphics[width=7.5cm]{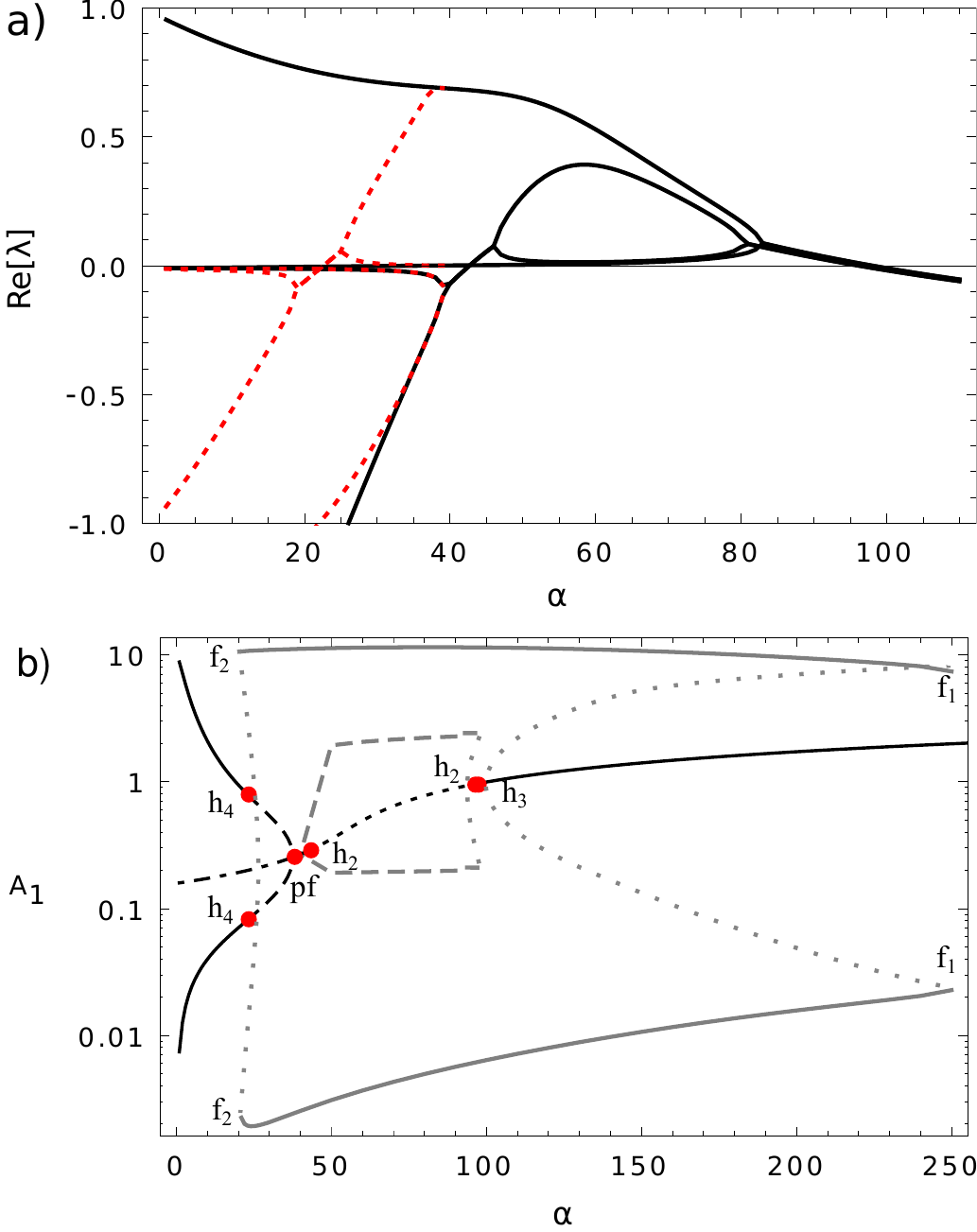}
	\end{center}
	\caption{(Color online) Bifurcation diagram for eigenvalues (upper panel) and fixed points and limit cycles (lower panel) as a function of $\alpha$ for $\beta=10$. Gray (red) dashed lines represent the real parts of the eigenvalues of the two non-homogeneous, symmetric fixed points, black full lines those of the homogeneous fixed point. In the lower panel black lines represent fixed points, full lines  stable ones,  otherwise unstable (dashdotted three directions stable and one unstable, dashed two stable and two unstable directions, dotted all four directions unstable). Gray lines represent limit cycles, full lines  stable ones, dashed saddle, and dotted unstable ones.}
\end{figure}

The regimes $FP1$ and $FP3$ are characterized by the same attractor, which is a single homogeneous stable
fixed point with positive values for the concentrations $A_i, B_i, i=1,2$, all of them having the same value at the stationary state. This fixed point first
looses and later gains stability through two Hopf bifurcations (indicated along the lines $h_1$ and $h_3$). The regime $FP1$ is surrounded by $\alpha = 0$ and $\beta = 0$-lines and a Hopf bifurcation branch ($h_1$). The branch $h_1$ starts at $\alpha= 0$, $\beta = 0.04$ with a  supercritical bifurcation, and at $(\alpha,\beta) = (16.21, 0.29)$ it becomes subcritical. The branch $h_3$ is always subcritical. The regime $FP3$ is bounded by a branch $f_1$, which is
a fold bifurcation of limit cycles, and the $\beta = 0$ line. This regime seems not to be bounded for $\alpha\rightarrow\infty$.\\
The regime $FP2$ is also a fixed-point regime. For this choice of parameters, there are three
fixed points, two stable and symmetric ones, and one unstable. At the unstable fixed point
 all variables have the same value, it is the one that is the attractor in the
regimes $FP1$ and $FP3$. In regime $FP2$ this fixed point has negative real parts of three eigenvalues and one real part being positive, see Figure~\ref{figure3}(a). For the two stable fixed points in this regime we have
$A_1 = B_1 = X \not= Y = A_2 = B_2$ or $A_1 = B_1 = Y \not= X = A_2 = B_2$. Therefore they are symmetric with respect to $A_1=B_1=A_2=B_2$, and have the same eigenvalues. The system evolves to
one of the two stable fixed points depending on the initial conditions. This regime is bounded
by the $\alpha = 0$-line and the fold bifurcation branch $f_2$.

\subsubsection{Regimes with coexisting fixed points and limit cycles}
$FL1$ and $FL2$ are regimes with three and two attractors, respectively. In the $FL1$-regime the attractors are the two stable fixed points from regime $FP2$ and a stable limit
cycle from regime $LC3$. The two stable fixed points loose their stability at the same time
through the subcritical Hopf bifurcations $h_4$. This leads to the creation of an unstable limit cycle,
which collides with a large stable limit cycle in a fold bifurcation $f_2$ of the two limit cycles,  where
both of them disappear. A similar scenario is observed in the $FL2$-regime. Here the unstable
fixed point from the $LC1$-regime gains stability through the subcritical Hopf bifurcation $h_3$, which
again causes the creation of an unstable limit cycle. This unstable limit cycle collides with a large stable limit cycle in a fold bifurcation $f_1$ of the limit cycles,  where both disappear and the system enters the $FP3$-regime.

\begin{figure}\label{figure4}
	\begin{center}
		\includegraphics[width=6cm]{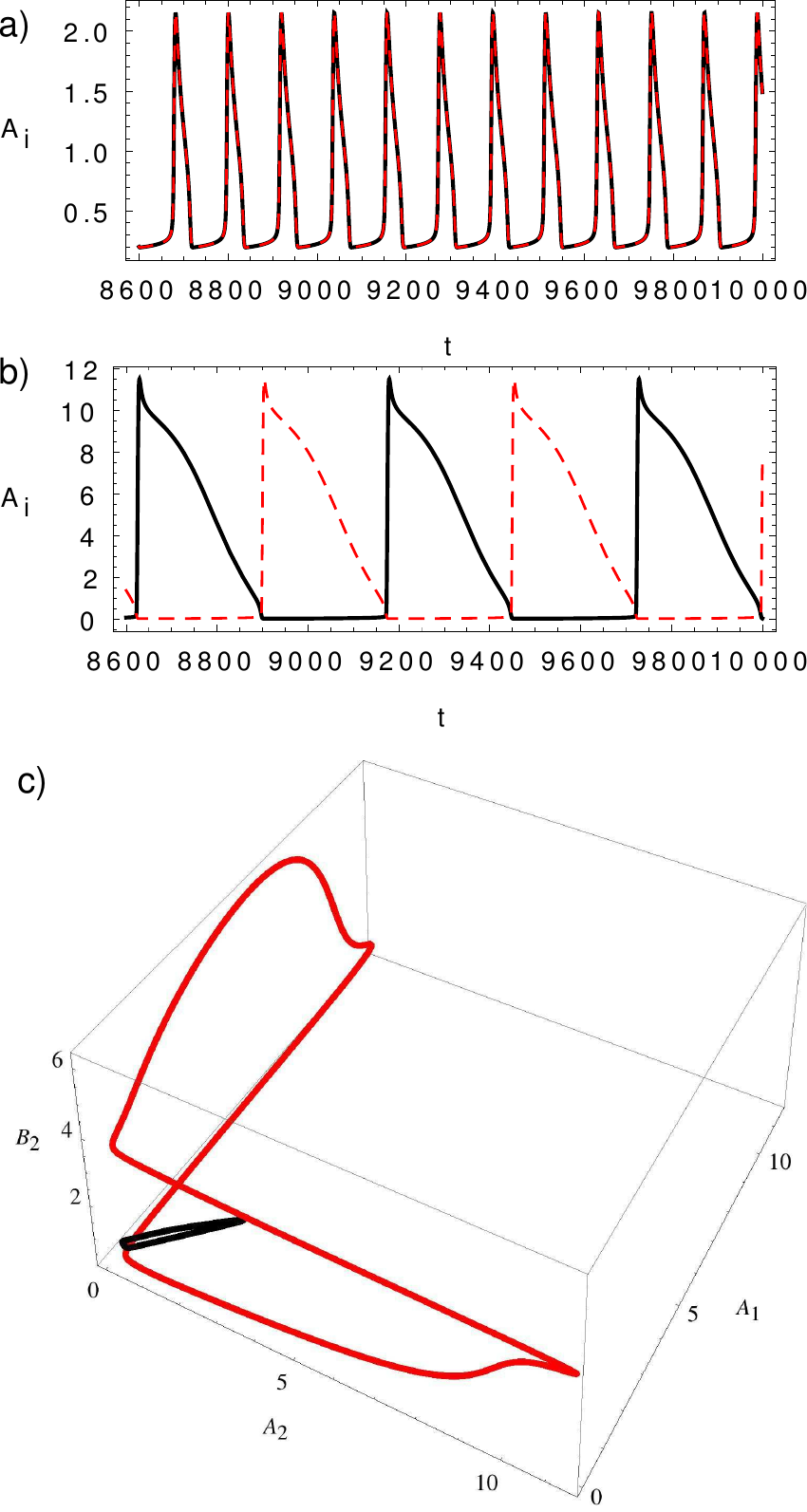}
	\end{center}
	\caption{(Color online) Time evolution of $A_1$ (black full line) and $A_2$ (gray (red) dashed line) and a three-dimensional phase diagram of $A_1$,  $A_2$ and $B_1$ for $\beta = 10$ and $\alpha=70$ for different initial conditions. The figure illustrates the difference between the two coexisting attractors. Figure (a) shows a saddle one-cluster and small-amplitude limit cycle, (b) a stable two-cluster and  large-amplitude limit cycle. The difference in the size is represented very well in a phase portrait in (c).}
\end{figure}

\subsubsection{Limit-cycle regimes}

\begin{figure}\label{figure5}
	\begin{center}
		\includegraphics[width=6cm]{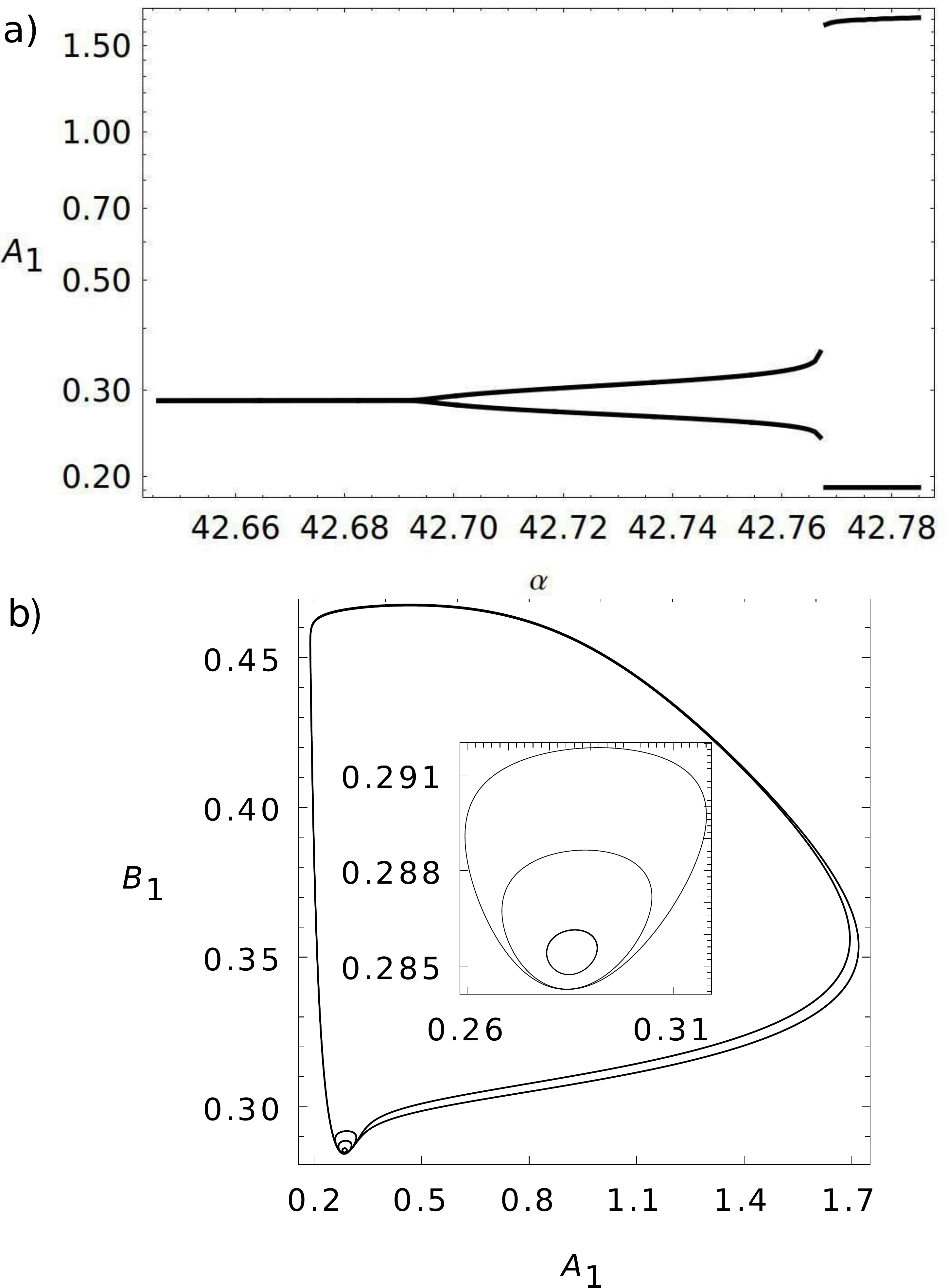}
	\end{center}
	\caption{Canard explosion from a small saddle-limit cycle to a large relaxation cycle for $\beta=10$. Figure (a) shows the growth of a limit cycle with an increase of $\alpha$, (b) a phase portrait of $A_1$ and $B_1$ for four values of $\alpha$ varying between $42.74-42.78$ in steps of $0.01$ from the smallest to the largest cycle, respectively. The inset in the second figure shows a zoom into   the first three cycles.}
\end{figure}

Regime $LC1$ is characterized by two coexisting limit cycles. In this regime there is only
one unstable fixed point, for which the real parts of all eigenvalues are positive. The regime is
bounded by the $\beta = 0$-line and the $h_2$-branch. For $\beta > 0.003$, one limit cycle is created through a supercritical Hopf bifurcation, when the fixed point looses its stability and a saddle-limit cycle is created. Before the Hopf bifurcation, the Jacobian at the fixed point  has eigenvalues with two positive and two negative real parts. At this bifurcation  also the two negative real parts become positive, and the fixed point looses its stability in all directions of the phase
space, see Figure~\ref{figure3}(a). The limit cycle created in this way is a saddle-limit cycle: Starting with the appropriate initial conditions, the system evolves to that limit cycle, but after a sufficiently long  time (of the order of 30000 t.u., corresponding to approximately 1000 periods), the system leaves the saddle-limit cycle and evolves
to a larger, completely stable limit cycle. (Such long-living transients like the saddle-limit cycle may be relevant in biological applications, where the transients matter if the system has no time to approach the attractors.) The bifurcation diagram in Figure~\ref{figure3}(b) in the interval $\alpha \in (40,100)$ corresponds to the $LC1$-regime, where the gray dashed line corresponds to the saddle-limit cycle.   These two type of  limit cycles can have very similar amplitudes, but they differ in their synchronization pattern,  so that they are easily distinguishable. The
saddle-limit cycle is a one-cluster limit cycle $(A_1(t) = A_2(t)$, and $B_1(t) = B_2(t))$, while
the stable limit cycle is a two-cluster limit cycle $(A_1(t) \not= A_2(t)$, and $B_1(t) \not= B_2(t))$, see Figure~\ref{figure4}. (The concentrations $A$ differ from $B$ in all limit-cycle regimes.) In addition, the saddle-limit cycle starts growing slowly with an increase of the bifurcation parameter as it is predicted for supercritical Hopf bifurcations. After a very small increase of $\alpha$, however, it goes through a canard explosion, characteristic for systems with slow and fast variables, see Figure~\ref{figure5}.

\begin{figure}\label{figure6}
	\begin{center}
		\includegraphics[width=7.5cm]{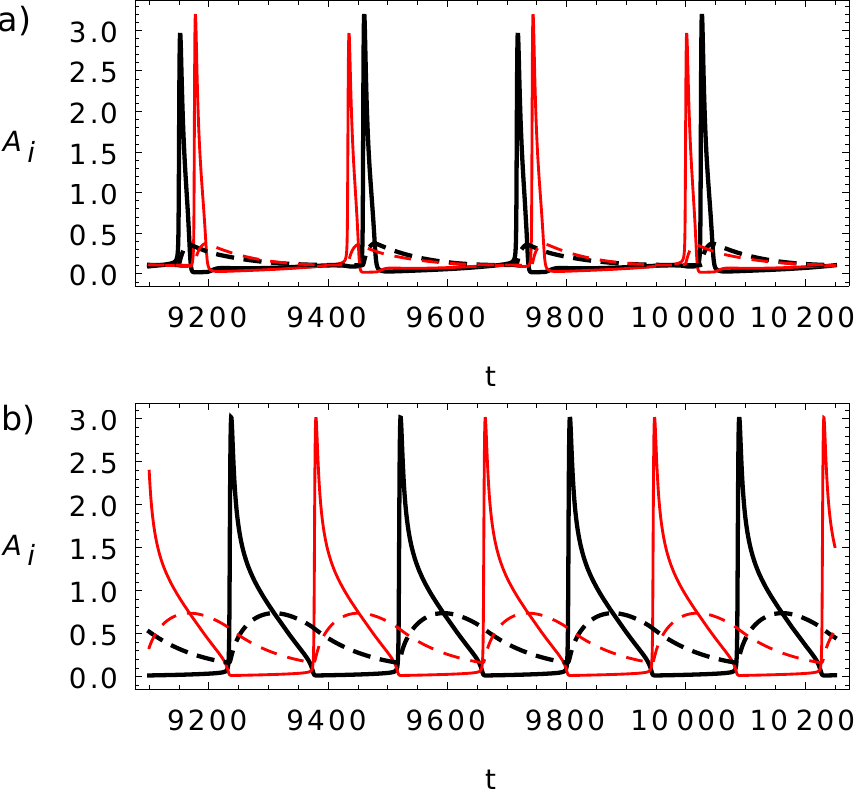}
	\end{center}
	\caption{(Color online) Time evolution of $A_1$ (black full line), $A_2$ (gray (red) full line), $B_1$ (black dashed line), and $B_2$ (gray (red) dashed line) for the regimes LC2 (first plot $\beta=0.1$) and LC3 (second plot $\beta=1$). The limit cycles differ in their frequency and periodicity. The parameters are $\alpha = 31.5$,  $\gamma = 0.01$, $b = 0.01$, and $K=0.02$.}
\end{figure}

For $\beta < 0.003$ all Hopf bifurcations are subcritical, so the stable limit cycles are created through  fold bifurcations of limit cycles. There are still two limit cycles as for $\beta>0.003$, one stable two-cluster limit cycle, and one saddle-one-cluster limit cycle, for more details see appendix~\ref{det_bif}.\\
\noindent In contrast to regime $LC1$, the regimes $LC2$ and $LC3$ are characterized by a single limit-cycle attractor. In both cases the stable
limit cycle is a relaxation cycle with a large amplitude. At the branches $h_1$ and $h_4$, an unstable limit cycle is created. This limit cycle collides with the relaxation cycle outside of the $LC2$ and $LC3$ regimes in a fold bifurcation of limit cycles. The limit cycles of the two regimes mainly differ in the oscillation periods of their phases (a two-period for $LC2$, and a one-period for $LC3$), see Figure~\ref{figure6} ((a) and (b), respectively.

\subsection{Bifurcation analysis along sections of constant $\beta$ for varying $\alpha$}\label{secbifurcation}
\begin{figure}\label{figure7}
	\begin{center}
		\includegraphics[width=8cm]{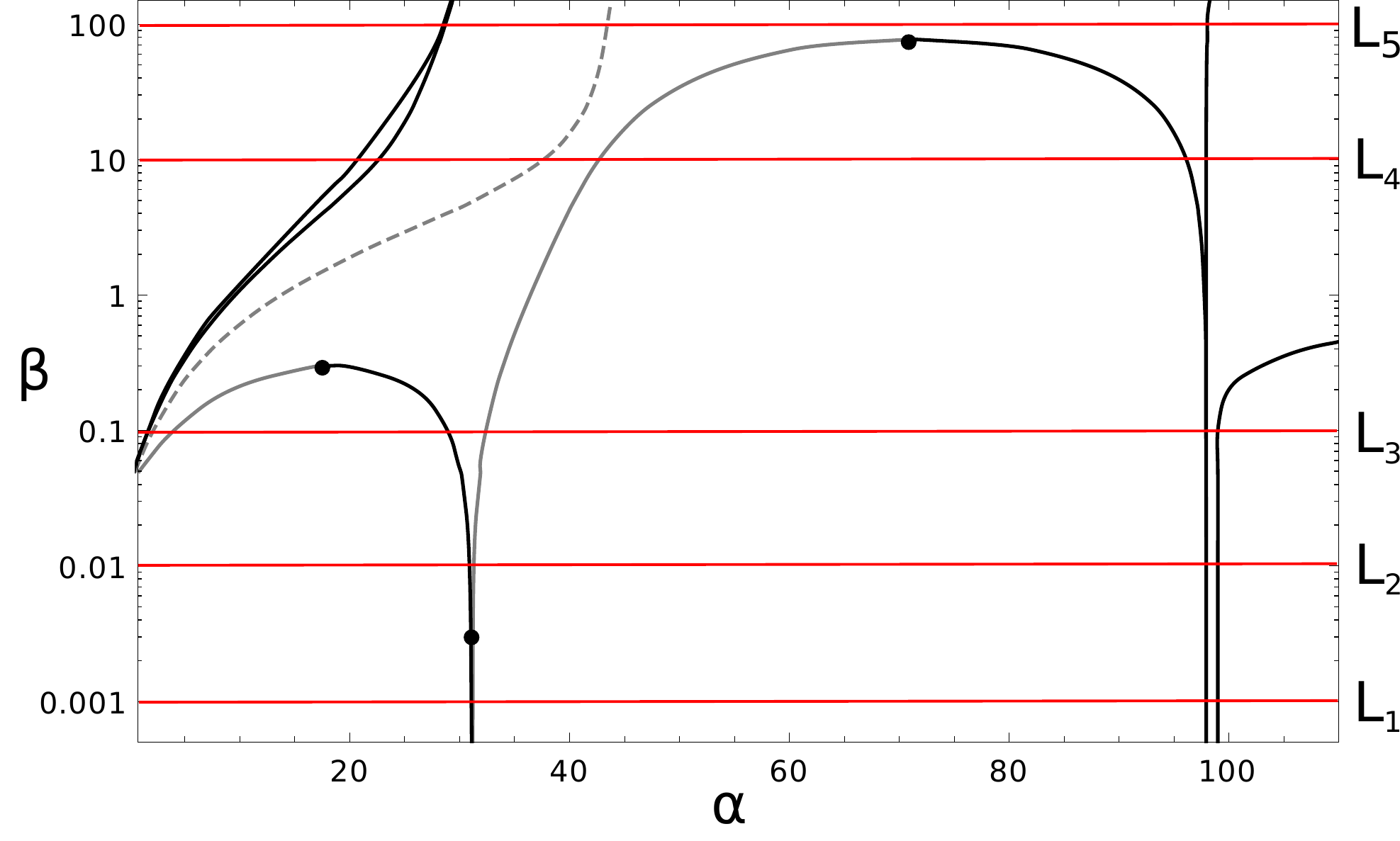}
	\end{center}
	\caption{(Color online) Bifurcation diagram as in Figure~\ref{figure2}. Gray (red) lines represent cross sections for constant $\beta$ as described in detail in the text and the appendix~\ref{det_bif}.}
\end{figure}

\begin{figure*}\label{figure8}
	\begin{center}
		\includegraphics[width=12cm]{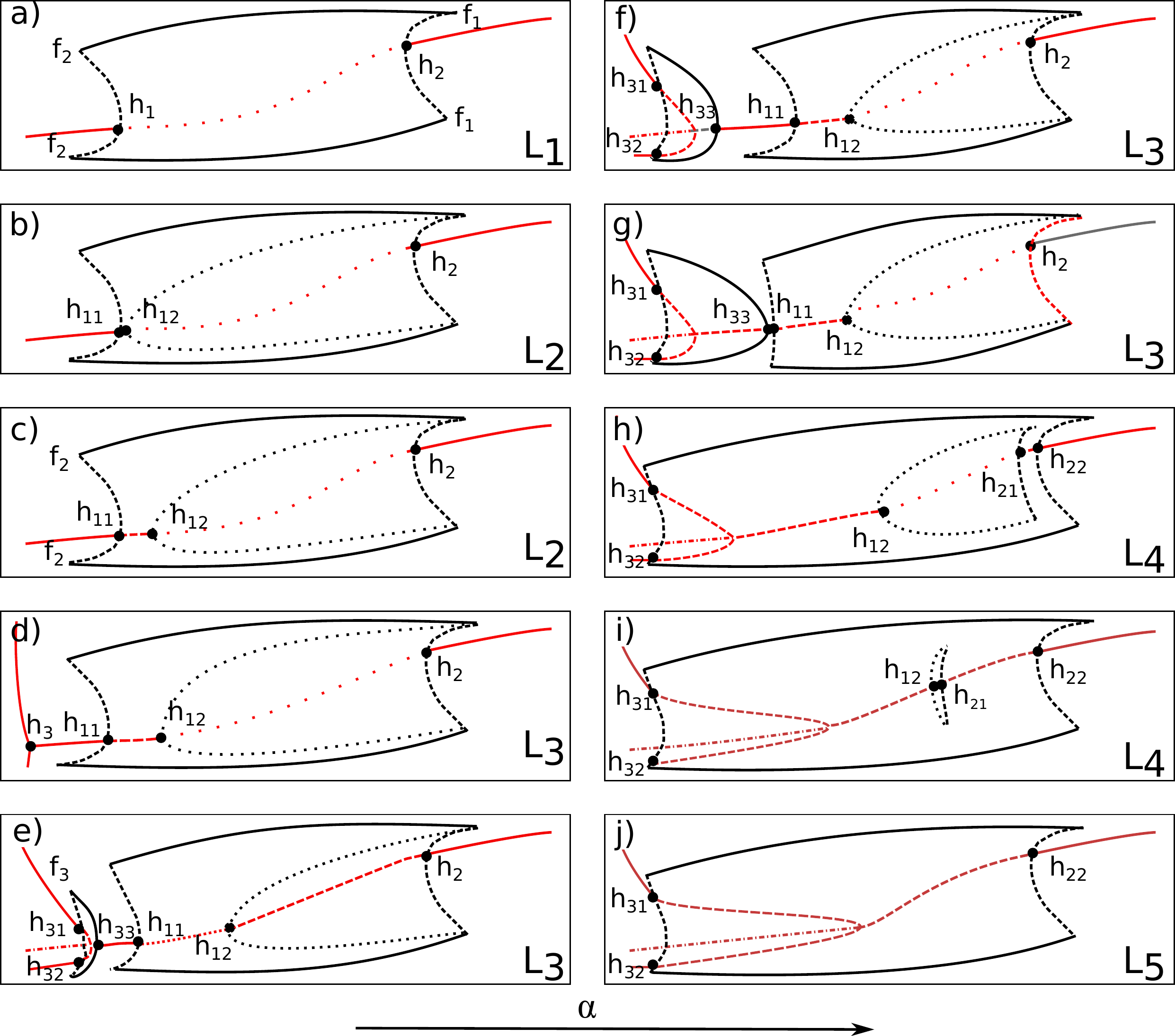}
	\end{center}
	\caption{(Color online) Sketches of the bifurcation diagram from $\beta=0$ to $\beta = 100$. Black lines represent the minimum and maximum values of the limit cycles, full lines represent stable limit cycles, dotted lines saddle-limit cycles, and dashed lines unstable limit cycles. Gray (red) lines represent fixed points, full gray (red) lines correspond to stable fixed points, dashdotted lines are fixed points with three stable directions and one unstable, dashed lines with two stable and unstable direction, and dotted lines represent fixed points with all four directions being unstable. Black points represent Hopf bifurcations $h$.}
\end{figure*}

It is instructive to pursue the bifurcations as a function of $\alpha$ for different fixed values of $\beta$ that are indicated as red (online color) lines in Figure~\ref{figure7}. We select five values of $\beta$ that are representative for different sequences of stationary states, which are approached when the respective regime of the bifurcation diagram is traversed with increasing $\alpha$.  While Figures~\ref{figure2} and \ref{figure7} are based on interpolated data, the representation along the panels of Figure~\ref{figure8} below is schematic to make the figure easier readable.

So Figure~\ref{figure8} zooms into the bifurcation diagram of Figure~\ref{figure7} with respect to $\alpha$ on the abscissa and for different, but fixed values of $\beta$, for which the corresponding panels are then representative for a whole interval around these $\beta$ values. On the ordinate we plot either fixed-point values (gray (red)), stable or unstable, or the minimal and maximal values of the limit cycles (black), stable or unstable, where the style of lines codes the stability properties as described in the caption.

For $\beta=0$, which corresponds to an uncoupled system with two identical units since all parameters are chosen to be the same, there are two subcritical Hopf bifurcations ($h_1$ and $h_2$), see Figure~\ref{figure8}(a). The system evolves to a limit cycle for $\alpha(h_1) < \alpha < \alpha(h_2)$ and to a homogeneous fixed point otherwise. Here $\alpha(h_i), i=1,2$ denotes the value of $\alpha$, at which the Hopf bifurcation $h_i$ happens. The limit cycles created in these subcritical Hopf bifurcations are unstable, they grow and collide with a stable limit cycle at the fold bifurcations of limit cycles ($f_1$ and $f_2$). The stable limit cycle exists for $\alpha(f_1) < \alpha < \alpha(f_2)$.

After coupling the system, this scenario holds for a small interval in $\beta$, more precisely for $0 < \beta < 0.003$, after which the $h_1$ branch divides into two branches ($h_{11}$ and $h_{12}$), one subcritical and one supercritical Hopf bifurcation branch, see Figure~\ref{figure8}(b). After the splitting of this Hopf bifurcation, an additional saddle-limit cycle is created by the supercritical Hopf bifurcation  $h_{12}$. For $\alpha \in (\alpha(h_{12}), \alpha(h_2))$ the system can evolve either to a large stable limit cycle or to a smaller saddle-limit cycle, depending on the initial conditions. After a rather long time the system will leave the saddle-limit cycle, and evolve to a stable limit cycle.

With increasing $\beta$, $h_{11}$ and $h_{12}$ separate from each other, $h_{11}$ moves to smaller values of $\alpha$ and $h_{12}$ to larger values, see Figure~\ref{figure8}(c).

At $\beta \sim 0.05$ two additional symmetric fixed points emerge through a pitchfork bifurcation, see Figure~\ref{figure8}(d). The homogeneous fixed point has one unstable direction, while the symmetric fixed points are stable, so for $\alpha < \alpha(h_3)$ the system evolves to one of the symmetric fixed points.

From the point $h_3$, three additional Hopf bifurcations  emerge: ($h_{31}$, $h_{32}$, and $h_{33}$), see Figure~\ref{figure8}(e), one at each fixed point. The Hopf bifurcations at the symmetric fixed points ($h_{31}$ and $h_{32}$) happen at the same value of $\alpha$, unstable limit cycles are created, which collide with a stable limit cycle at the fold bifurcation of limit cycles $f_3$. The Hopf bifurcation at the homogeneous fixed point $h_{33}$ is supercritical, and the stable limit cycle, which is created at that point, is the limit cycle that collides with the unstable limit cycles created by $h_{31}$ and $h_{32}$, see Figure~\ref{figure8}(e).

The Hopf bifurcations $h_{31}$, $h_{32}$ and $h_{33}$ further separate  with increasing $\beta$, as well as $h_{11}$ and $h_{12}$,  Figure~\ref{figure8}(f), but $h_{33}$ and $h_{11}$ approach each other until they collide, see  Figure~\ref{figure8}(g).
After the collision of $h_{33}$ and $h_{11}$, one stable and one unstable limit cycle are gone, neither exists an intermediate stable fixed-point regime between two limit-cycle regimes. The branch $h_2$ also splits into $h_{21}$ and $h_{22}$,  Figure~\ref{figure8}(h), and the two branches separate further with increasing $\beta$.

Figure~\ref{figure8}(i) shows the instant just before the collision of $h_{12}$ and $h_{21}$. The saddle-limit cycle that was surrounded by a large stable limit cycle  disappears, leaving the system with a similar dynamics as for very weak coupling, see Figure~\ref{figure8}(j): For small values of $\alpha$ the system evolves to one of the symmetric fixed points, for intermediate $\alpha$ to a limit cycle, and for large $\alpha$ to a homogeneous fixed point.

\begin{figure}\label{figure_asym}
	\begin{center}
		\includegraphics[width=8cm]{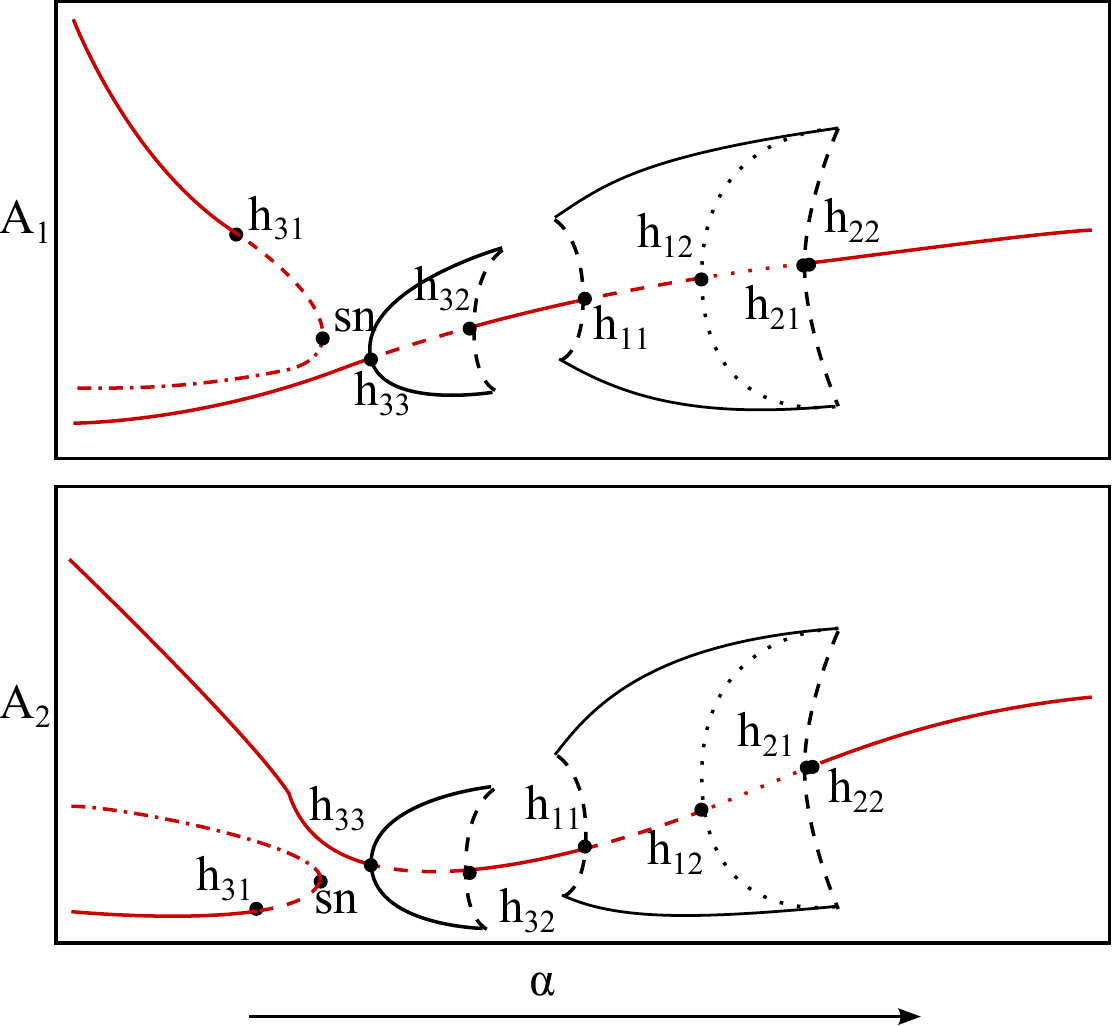}
	\end{center}
	\caption{(Color online) Sketch of the bifurcation diagram for two asymmetrically coupled  units as a function of $\alpha$. The couplings are chosen as $\beta1 = 0.2$ and $\beta2 = 0.21$. Gray (red) lines represent fixed points, black lines limit cycles and black dots bifurcations (as in Figure~\ref{figure8}). The bifurcation diagrams now differ for $A_1$ and $A_2$. The former pitchfork bifurcation has become a saddle-node bifurcation (sn), the Hopf bifurcation $h_3$ (called $h_4$ in Figure~\ref{figure2}) splits differently as compared to the symmetric case. Essentially $h_{31}$ and $h_{32}$ occur no longer simultaneously. The displayed  scenario should be compared with Figure~\ref{figure8} (f) of the symmetric case.}
\end{figure}

\vskip5pt
\noindent {\it Asymmetric coupling} So far our analysis applies for a system with symmetric coupling between both subsystems, given by $\beta$. It is of interest how an asymmetric choice of $\beta_1$ in one direction and $\beta_2$ in the reverse direction affects the bifurcation diagram. It differs in three aspects from the symmetric case.
Firstly, the amplitudes of the two units differ in the oscillatory regimes, and the fixed-point values in the fixed-point regimes. The difference is the larger, the larger the difference between the couplings. Secondly, for $\beta$-values, for which we encounter pitchfork bifurcations in the symmetric case (Figure~\ref{figure8} (e)-(j)), the pitchfork bifurcations are replaced by saddle-node bifurcations. The larger the difference  in the couplings, the more the two fixed points, which collide in the saddle-node bifurcation, separate from the third fixed point that exists for all positive $\alpha$. For the third difference  first recall that in the symmetric case the Hopf bifurcation $h_3$ of panel (d) splits up into three bifurcations $h_{31},h_{32},h_{33}$ in panel (e), where $\alpha(h_{31})=\alpha(h_{32})\not=\alpha(h_{33})$. In the asymmetric case the bifurcations still happen at two fixed points, but $\alpha(h_{31})\not=\alpha(h_{32})$. The difference in the corresponding $\alpha$-values increases with the asymmetry in the couplings. The bifurcation $h_{31}$
happens at the fixed point which disappears for larger $\alpha$, when it collides with the other unstable fixed point in a saddle node bifurcation, indicated with $sn$ in Figure~\ref{figure_asym}. It is subcritical, the limit cycle created there is unstable, and it is not detected in the numerical integration of the system. Instead, the system either evolves to a coexisting stable fixed point if $\alpha(h_{31})<\alpha(h_{33})$, or to a stable limit cycle if $\alpha(h_{33})<\alpha(h_{31})$, where $\alpha(h_{32})> \max\{\alpha(h_{31}), \alpha(h_{33})\}$. Therefore Figure~\ref{figure_asym} shows the case, where the system evolves to a stable fixed point. For the values of $\beta$ that we have checked ($\beta\in(0.05,20)$), the Hopf bifurcation $h_{33}$ is supercritical, the limit cycle created there is the stable two-cluster limit cycle.
In summary, in the asymmetric case the bifurcation diagram differs from the symmetric case, but as far as we have checked, the number of oscillatory regimes stays the same.

The bifurcation diagram was established by a combination of the program package MATCONT, MATHEMATICA  and code written in Fortran 90. Further details can be found in appendix~\ref{methods}.


\subsection{Functioning as a toggle switch}\label{nano}
Usually the toggle switch is considered between two mutually repressing genes that control each other's expression and exhibit bistability, so that the system can toggle between two discrete, alternative stable steady states. Such a toggle switch was one of the first synthetic gene networks that was constructed in Escherichia Coli \cite{gardner}. Gardner {\it et al.} in \cite{gardner} predicted the conditions which are necessary for bistability. In general the bistability can depend on the type of coupling (cooperative coupling or with competitive binding in an exclusive-or switch), or on the stochastic (versus deterministic) realization \cite{walczak}. Toggle switches in combination with an intercell communication system that mediates the interaction between the different cells were considered as basic building blocks of a whole population of genetic oscillators \cite{nancy}. Here we consider a single toggle switch, but itself being composed of more complex units than usually considered, namely the two mutually repressing circuits. Our system then provides a device that can switch not just between two, but between a multitude of states when the repressive coupling $\beta$ or the bifurcation parameter $\alpha$ are accordingly tuned. Keeping these parameters fixed, we have still two regimes in which the system can toggle between three (FL1 regime) or two (FL2 regime) stable states, when noise is added to the deterministic description. In view of biological applications, a stochastic description is more realistic to account for the inherent fluctuations of various origin, in particular the finite number of involved genes, proteins, or cells.

\subsubsection{Noise-induced switches in the bistable regime}

\begin{figure}[ht]\label{figure10}
\center
\includegraphics[width=7 cm]{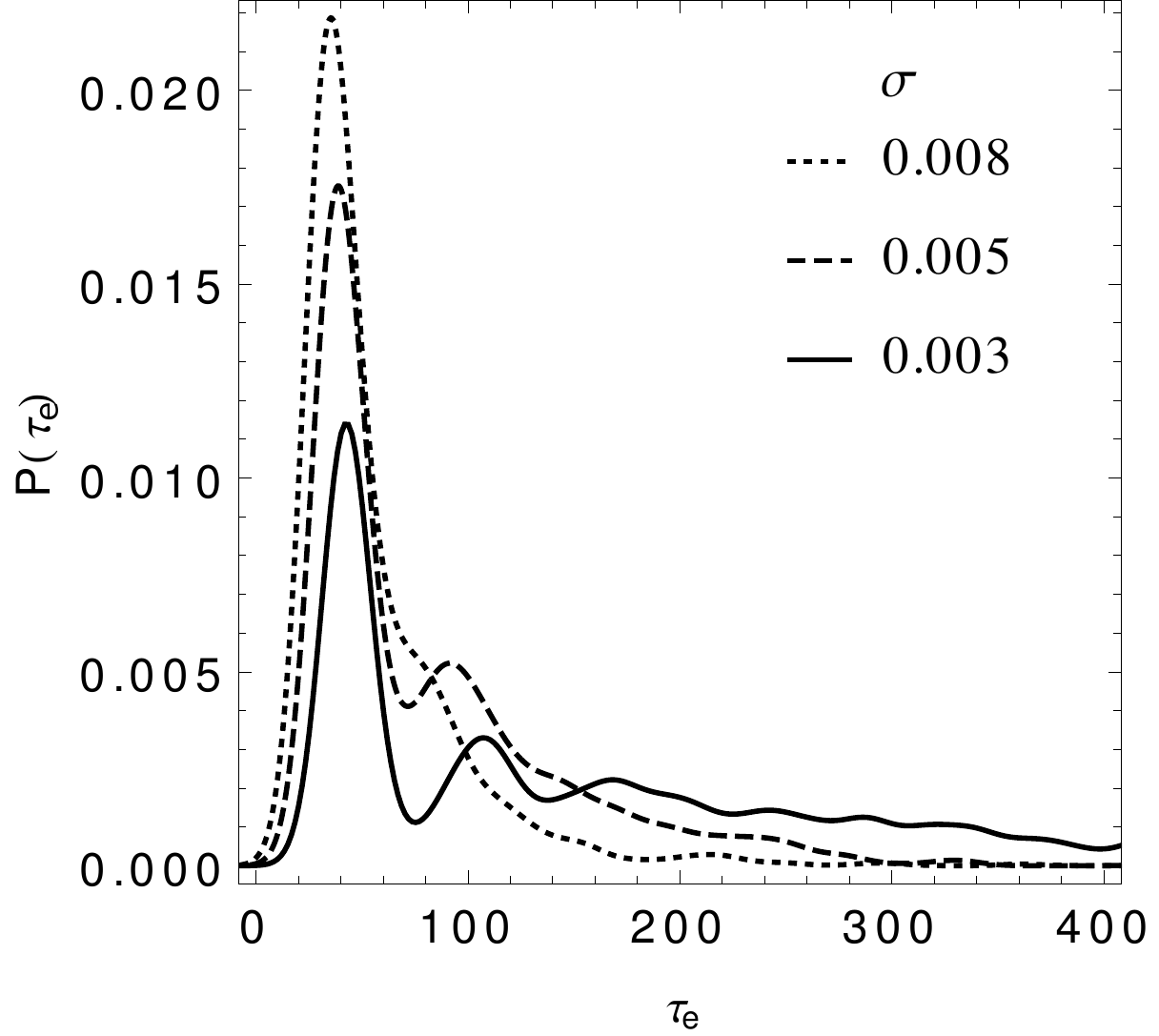}
\caption{Probability distribution of the escape times from a fixed point to a stable limit cycle through an unstable limit cycle under the influence of noise in the $FL2$ regime. The parameters are $K=0.02$, $b=\gamma =0.01$, $\alpha = 102$, $\beta = 1$, the noise intensity is chosen as $\sigma = 0.003$ (full line), $0.005$ (dashed), and $0.008$ (dotted).} 
\end{figure}

We study the switching between coexisting states in the multistable regime $FL2$ under the influence of Gaussian noise. We add a noise term to each of the equations  (\ref{eq3}) and measure the escape time from the fixed-point regime to a limit-cycle regime via the time of arrival at the large limit-cycle attractor, once the system left the fixed-point attractor. We solve the system of stochastic differential equations with the Heun method, adjusting the noise intensity $\sigma$ by factor $\sqrt{\Delta t}$ according to \cite{greiner}, where $\Delta t$ denotes the size of an integration step. We use the same noise intensity $\sigma$ for the four equations.
Figure~\ref{figure10} shows the probability distribution of the escape times for different values of the noise intensity $\sigma$ from the fixed point  to the stable limit cycle. These are the coexisting attractors in the $FL2$ regime, in which the escape proceeds via an unstable limit cycle. One can see that the probability distribution has a multi-peak structure. For weaker noise intensity (black full line in Figure~\ref{figure10}), multiple peaks are more pronounced than for stronger noise (black dotted line in Figure~\ref{figure10}). The maximum peak is around $\tau_e=50$, and it increases with an increase of the noise intensity, while the secondary maxima decrease or vanish completely. The multipeak structure is expected, because the escape happens via an unstable limit cycle, so that the trajectory may follow the unstable limit cycle once or twice or more often, before it approaches the stable limit cycle \cite{kaw}.

For $\alpha$ close to the subcritical Hopf bifurcation ($h_{22}$ in Figure~\ref{figure8}(h)), when the unstable limit cycle has a small radius and is very close to the stable fixed point, a very weak noise intensity will lead to an escape from the fixed point to a stable limit cycle, which surrounds both the unstable limit cycle and the stable fixed point. After the escape, there is a low probability for the system to return to the fixed point. On the other hand, if $\alpha$ is close to a fold bifurcation $f_1$, the radius of the unstable and stable limit cycles are similar. A very strong noise intensity is then needed for the system to escape from the fixed point to the remote stable limit cycle, and it will return to the fixed point with high probability. Conversely, a transition from the limit cycle to the fixed point is very likely, since a small noise may kick the system across the unstable limit cycle into the basin of attraction of the stable fixed point. For $\alpha$ away from the bifurcations (but in the FL1 regime) and for sufficiently strong noise, the system switches back and forth between the two states. The stronger the noise, the more frequent the switching. These expectations were confirmed by our numerical simulations of time evolutions for different noise intensities.


\subsubsection{Controlling the function via tuning the bifurcation parameters}

\begin{figure}\label{figure11}
\center
\includegraphics[width=8 cm]{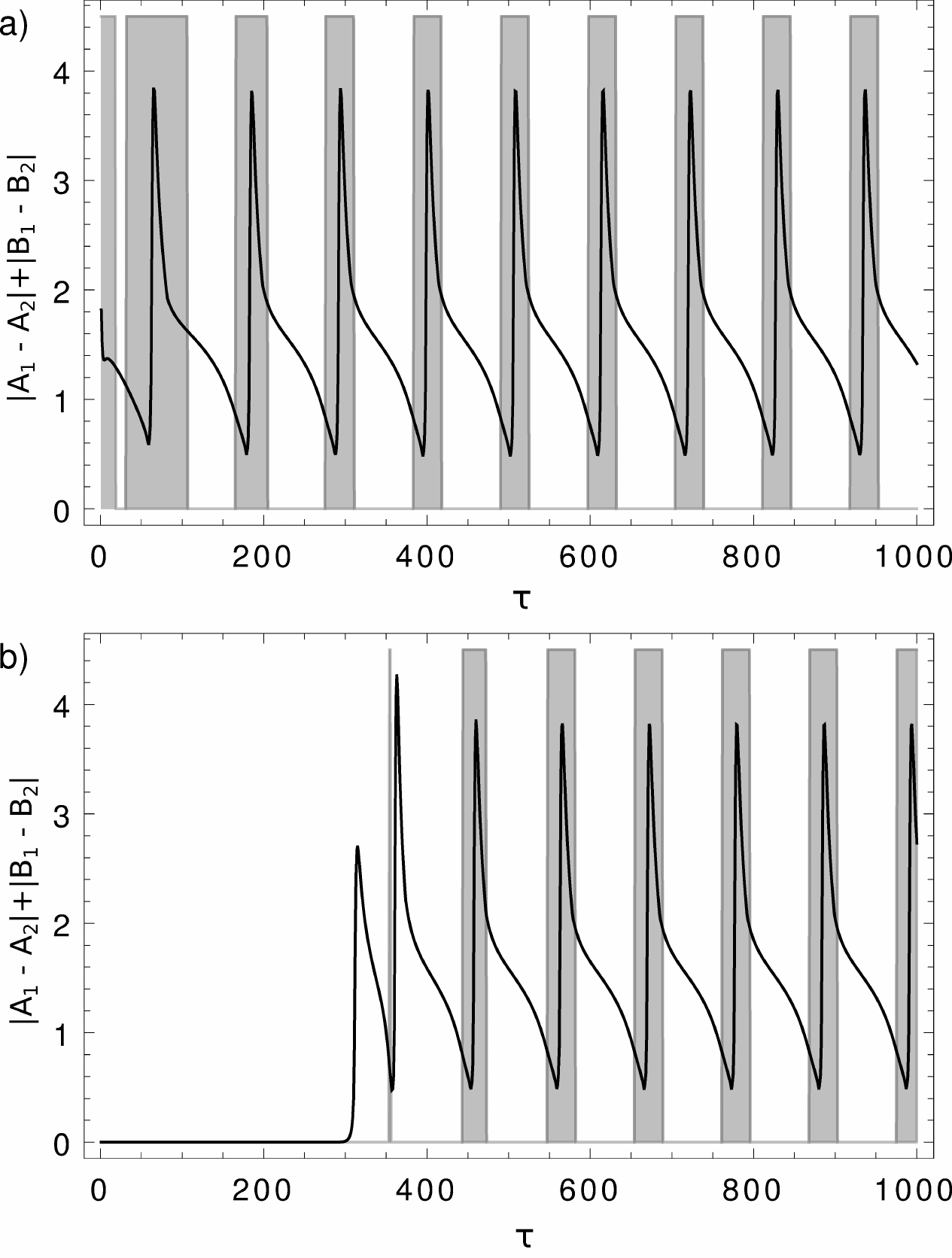}
\caption{Time evolution of a representative combination of the variables in the $LC1$ regime. Light shaded intervals indicate that the system will evolve to a limit cycle after it spent time $\tau$  in regime $LC1$. The figures show the results for two different initial states, a limit cycle (a) and a fixed point (b). The parameters are $K=0.02$, $\beta=\gamma=0.01$, $\beta=1$, and $\alpha = 110$ in the first and last $2000$ time units, while $\alpha = 60$ over the time span  $\tau$. The initial conditions are $A_1=B_1=1.1$, $A_2=B_2=0$ (a)  and $A_1=A_2=B_1=B_2=1.1$ (b).}
\end{figure}

We see various sequences of limit-cycle and fixed-point behavior. They possibly differ by the synchronization pattern of phases, the amplitude of oscillations, or the fixed point values.
Which sequence is seen depends on the path in the $(\alpha,\beta)$-space of Figure~\ref{figure2}, on the speed, by which the parameters are changed, on the dwell time, which the system spends in a certain regime.   In a bistable regime with coexisting states it is of particular interest how we can control the return to the initial state after a closed path in parameter space has been performed. For simplicity we keep $\beta$ fixed ($\beta=1$) and quench the system only in $\alpha$ from $110$ to $60$, that is from the bistable regime $FL2$ into the monostable regime $LC1$, let it evolve there for  certain time, and then quench it back from $60$ to $110$, back to $FL2$. Regardless of the initial state in $FL2$, the system always evolves to a stable two-cluster limit cycle in $LC1$. We then analyze the return into the $FL2$ regime as a function of the dwell time $\tau$ which the system spends in $LC1$. As it turns out, it is the values of $A_i$ and $B_i$  just before the quench back into $FL2$ that determine whether the system moves back to the initial state or to the alternative state of $FL2$. Obviously these values depend on the type of states which are available in $LC1$, and the dwell time which the system has to evolve there. For a regime like $LC1$ with regular oscillations  it is therefore not surprising that we see a periodic return to the initial state in $FL2$ as a function of the dwell time, as seen in Figures~\ref{figure11}.\\
Figures~\ref{figure11}(a) and (b) show the combined  values of $A_i$ and $B_i$, reached before the quench back to the initial regime as a function of $\tau$. The superimposed shaded stripes indicate whether the system returns to the fixed point (white stripes) or the limit cycle (gray stripes). For a short dwell time of the order of half a period, the system always returns to the initial state, while for times larger than one limit cycle period, the system periodically evolves to a fixed point or a limit cycle after the quench, if the amplitude of the limit cycle in LC1 is sufficiently large.

Other examples of quenches between LC3 and LC2 for small $\beta$ have shown that the system returned always to the fixed point, independently of the start in the bistable regime. The reason is that the amplitudes are not large enough to provide initial conditions at the quench back which could reach the basin of attraction of the limit cycle state in  $FL2$.

\section{Collective coherent behavior of a large number of coupled genetic circuits}\label{sec4}
For a single genetic circuit we  found three regimes with qualitatively different behavior: fixed-point behavior for small and large $\alpha$, differing by the fixed-point values, and limit-cycle behavior for an intermediate range of $\alpha$. For two coupled circuits we found already eight regimes, as described in the previous section. In a naive extrapolation one may expect a proliferation of regimes in larger sets of coupled circuits. So it may come as a surprise that for $50\times 50$-sets of coupled units we observe again a relatively simple structure of mainly three regimes, as for a single unit. This observation holds for a relatively broad range of uniformly chosen parameters $\alpha$ and $\beta$. The larger system can be neither addressed with a detailed bifurcation analysis as in the previous section, nor is a simple  extrapolation of the number of distinct regimes towards a larger system size accessible. Already for a system of four units it is easily seen in appendix~\ref{4units} that the previous Hopf bifurcations at small and large $\alpha$ split into three or four bifurcation branches if $\beta>0$. The multistability in certain regimes becomes more pronounced. In contrast, for sufficiently large $\beta$, branches of Hopf bifurcations may disappear, so that the number of distinguishable regimes shrinks again to a few. Therefore our investigations are on a mere phenomenological level and limited to numerical simulations, however, with possible very interesting applications to biological systems. The reason is
that it is the single parameter $\alpha$, whose monotonic increase allows to turn on the collective oscillations in a first bifurcation and to arrest these oscillations later in a second bifurcation. Its tuning therefore provides a mechanism for controlling the duration of collective oscillations in a simple way, without external interference and without finetuning of parameters. Because of its simplicity, such a mechanism may be realized in biological applications. In particular it may be relevant for the segmentation clock.

The segmentation clock stands for an oscillating multicellular genetic network that controls the sequential subdivision of the elongating vertebrate embryonic body
axis into morphological somites (these are regularly sized cell clusters). The morphogenetic rhythm of the somitogenesis is based on repeated waves of oscillatory gene expression sweeping through the tissue. Usually it is the clock and wavefront mechanism \cite{1976} that is supposed to translate  the oscillations in time of a clock into a periodic pattern in space. It happens when a wavefront moves posteriorly  across the presomitic mesoderm (PSM) from the anterior and freezes the cellular oscillators as it passes by, recording their phases at the time of arrest. Questions of interest  concern the relation between the patterns on the genetic and the cellular levels and the mechanism how a collective segmentation period rises from the ensemble of individual units at all.

In \cite{juelicher} a delayed coupling theory was developed as a phenomenological  mesoscopic description of the vertebrate segmentation in terms of phase oscillators to represent cyclic gene expression in the cells of the PSM. The oscillators are coupled  to their neighbors with delay and have a moving boundary, describing the elongation of the embryo axis. They are equipped with a profile of natural frequencies that is particularly designed to slow down the oscillations and stop them at the arrest front.
The very mechanism of arresting the oscillations at the arrest front is, however, not addressed, neither is the onset of oscillations described as a Hopf bifurcation from an underlying dynamical system (as, for example, in \cite{jensen2003} or in \cite{goldbeter2008}). Differently, in \cite{goldbeter2007} and \cite{santillan2008}, the arrest of oscillations is supposed to result from an external signal (rather than from a suitably chosen frequency distribution or a system-inherent bifurcation).

As we shall see in the following, in our model the oscillations get arrested merely due to a second bifurcation within one and the same model.
As dynamical equations for our set of $50\times 50$ circuits we choose the Eqs.~\ref{eq3} with a uniform choice of  parameters $K$, $b$ and $\gamma$, using the same values as introduced for a single unit, that is $K=0.02$, $b = \gamma=0.01$. The parameter $\alpha$ is also chosen independently of the lattice site, but as a bifurcation parameter it is varied between $1$ and $110$ or $400$, depending on the repressive coupling strength, parameterized by $\beta$, being weak ($\beta=0.1$), or strong ($\beta=10.0$), respectively. Also $\beta$ is chosen with the same value on all bonds and in opposite directions.
We choose free (f.b.c.) or periodic (p.b.c.) boundary conditions. For p.b.c. the lattice sites in both directions are identified modulo L, the linear extension.
In order to facilitate the recognition of patterns we choose regular network topologies.

\begin{figure}\label{figure14}
	\begin{center}
		\includegraphics[width=6cm]{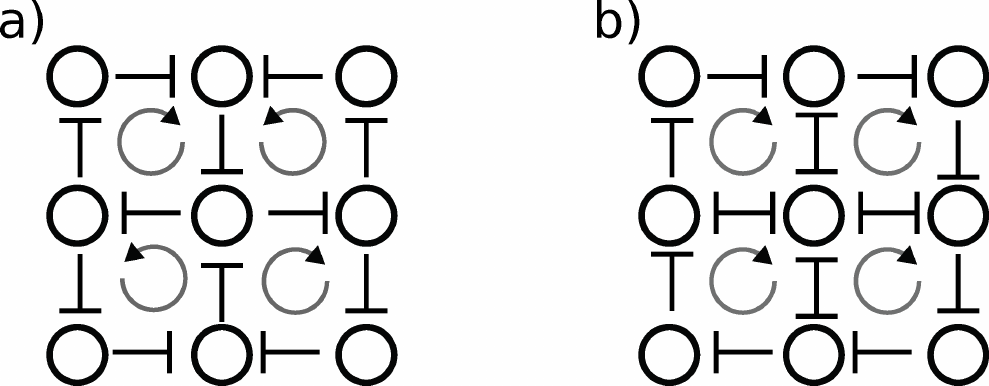}
	\end{center}
	\caption{Topologies (a) OC-lattice, (b) SC-lattice. At each vertex (circle) we have one circuit repressing two, three, or four neighbors. Blunt arrows represent repressive coupling, while pointed circular arrows represent the ``flow" of repression in the smallest possible loop on the lattice. If two such neighboring loops have a circulation of repression in opposite directions, we say that the lattice has opposite chirality (a), otherwise the chirality is the same (b).}
\end{figure}

From the various choices for the adjacency matrix $R_{ij}$ which we have studied in \cite{arxivdlhmo}, we show here just two  as indicated in Figure~\ref{figure14}:  square lattices of size $L\times L$ with $L=50$ which differ by the relative chirality of adjacent plaquettes.
In the topology of Figure~\ref{figure14}(a) adjacent plaquettes have opposite chirality.  In the alternative lattice of Figure~\ref{figure14}(b) with f.b.c., we choose the bonds along the boundary to be directed, but undirected in the bulk. So the plaquettes have the same chirality when each bond in the bulk is traversed twice but in opposite directions. We shall refer to the lattice in Figure~\ref{figure14}(a) as OC-lattice, and to the lattice in Figure~\ref{figure14}(b) as SC-lattice according to their chirality. The apparently minor features of the chirality turn out to be relevant for the specific patterns and their sensitivity to the boundary conditions; the three regimes which are traversed are nevertheless independent of this choice.

In view of illustrating the control mechanism via varying the parameter $\alpha$, we have  furthermore to specify the speed by which $\alpha$ is varied. An obvious choice, which is, however, not necessary  in view of biological applications, is an adiabatic change such that the system has time to approach one of the attractors before $\alpha$ is further increased.

In the stationary state, for small and large values of $\alpha$, we have collective fixed-point behavior (CFP), in which all individuals approach a fixed point in the concentration, though not necessarily of the same value. The collective oscillations (CO) are synchronized in frequency, but can differ by their patterns of synchronized phases.  So the collective fixed point and oscillatory behavior can be further characterized by the number of clusters, if we collect all units $i$ for which $A_i$ ($B_i$) agree within the numerical accuracy in one cluster.

\begin{figure}\label{figure15}
	\begin{center}
		\includegraphics[width=8.5cm]{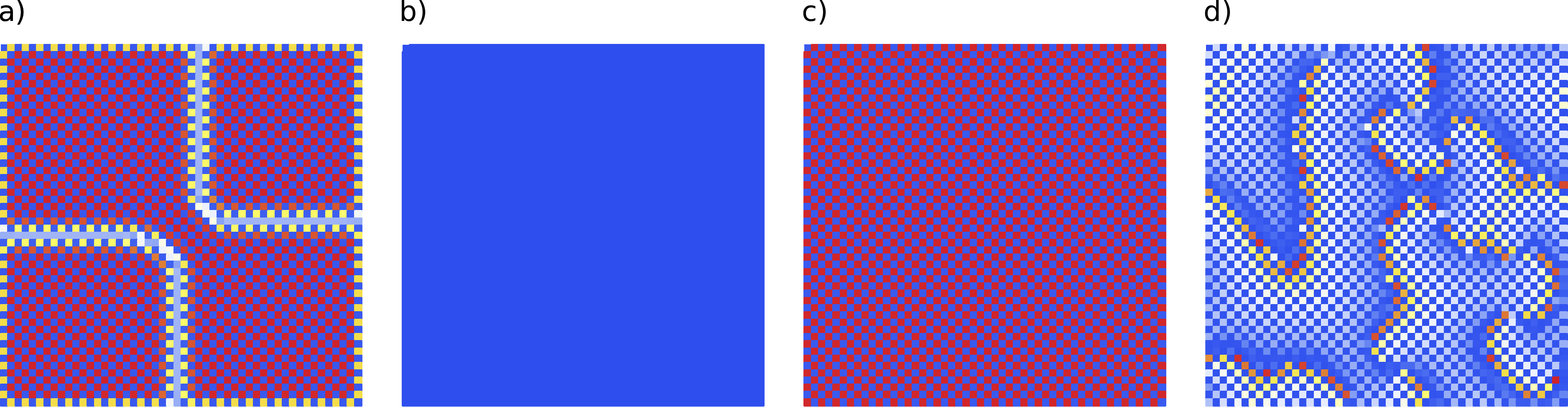}
	\end{center}
	\caption{Examples for stationary states for fixed $\alpha$ and weak coupling. Multi-cluster fixed point (a), one-cluster fixed point (b), two-cluster limit cycle (c) and multi-cluster limit cycle (d). For further explanations see the text.}
\end{figure}

Figure~\ref{figure15} shows snapshots of the values of $A$ for two topologies: SC-lattice (a), and OC-lattice (b)-(d). Figures~\ref{figure15}(a) and (b) show  fixed points, multi-cluster and one-cluster, respectively, where we started with slightly perturbed  uniform (a) and random (b) initial conditions. Figures~\ref{figure15}(c) and (d) show snapshots of two different limit cycles for the OC-lattice, two-cluster and multi-cluster limit cycles, respectively.

In general, for the $50\times 50$ lattices and fixed $\alpha$, the transient time to approach a fixed point is of the order of some hundred to thousand time units, while it can take up to 200000 time units to approach the synchronized limit cycles.

\begin{figure}\label{figure16}
	\begin{center}
		\includegraphics[width=8.5cm]{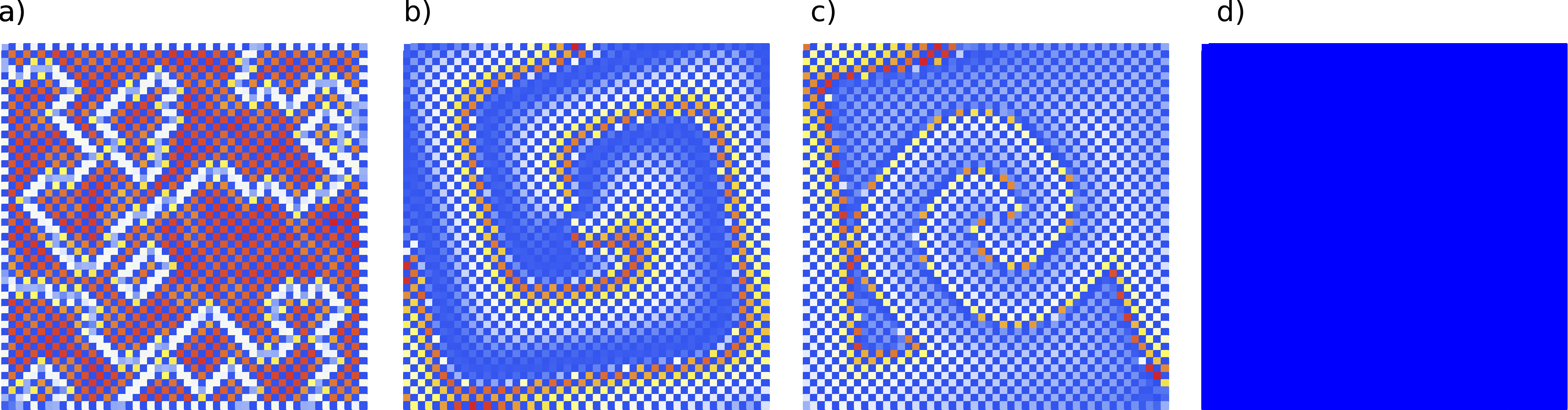}
	\end{center}
	\caption{Snapshots of collective fixed-point behavior (a), two kinds of collective oscillatory behavior (b) and (c), and again collective fixed-point behavior (d). Simulations are done for a gradual variation of $\alpha$, increasing from left to right in the panels, and weak coupling.}
\end{figure}

If we do not await the stationary states, but change $\alpha$ faster, we see transient patterns, however, the overall collective behavior remains the same: a relatively fast approach of the fixed point values and transient synchronized oscillations. These are shown in Figure~\ref{figure16}.

Figure~\ref{figure16} shows snapshots of the values of $A$ for the SC-lattice with f.b.c. and  at time instants from the CFP-regime (a) and (d), and from the CO-regime (b) and (c). All simulations were run for a gradual variation of $\alpha$, that is, $\alpha$ is changed by $0.0075$ every time unit, and at weak coupling $\beta=0.1$, while the other parameters were chosen as usual ($K=0.02$, $b=\gamma=0.01$). The color codes the value of $A$: ranging from blue for a minimal value to red for a maximal value. The snapshots were taken  after  100, 6000, 13000, and 15000 t.u., respectively. In the latter case, the spiral patterns created in the transient phase are preserved as stationary states.

So apart from a few exceptions, the sequence of CFP-CO-CFP is seen for various lattice topologies, various values of the variation speed and couplings, and for a broad range of parameters. For the exceptions we refer to \cite{arxivdlhmo}.

In contrast, the very type and duration of collective transient patterns depends on the network topology, the boundary and initial conditions and the strength of the repressive coupling. As initial conditions we have usually  chosen either random or slightly perturbed ordered initial conditions. While $\alpha$ is changed, the large variety of transient patterns in an intermediate range of $\alpha$ results from the variety of structured patterns, which evolve over a certain amount of time as long as $\alpha$ is kept fixed, and then serve as new initial conditions when a new value of $\alpha$ is chosen. How much of the rich transient behavior can actually unfold depends on the speed of variation of the bifurcation parameter.

The system (\ref{eq3}) is integrated with the fourth-order Runge-Kutta algorithm with an integration step size of $dt=0.01$, which turned out to be sufficiently small.

\subsection{Relevance for the segmentation clock}
If we specialize the lattice to effectively one-dimensional sets of size $1000\times 2$, we see stripes rather than spirals as typical patterns similar to the transient patterns as they occur in the segmentation clock. Such geometries better reflect the elongating embryos while the segmentation clock is turned on.
The question remains to what extent our model of coupled circuits may match the segmentation clock on an effective mesoscopic level of description. On such a level the interacting cells are described in terms of our model, arising from coupled positive and negative feedback loops on the underlying genetic level.
Our mechanism of pattern generation is different from the clock and wavefront mechanism \cite{1976}. We can create periodic patterns in space due to the traveling waves in the oscillatory period, or as a collective  multi-cluster fixed point solution of the circuits in the small-$\alpha$-regime. The multi-cluster fixed points are characterized by coexisting different, but frozen values of the concentrations $A$ and $B$. Due to the repulsive coupling no neighboring oscillators share the same value, unless the individual dynamics dominates the coupling term. Therefore the question arises whether there is any experimental support for our type of mechanism being at work in the segmentation clock. As explained before, given a system of coupled circuits like ours, each composed of a negative and positive feedback loop, the driving force for the onset and arrest of oscillations is the variation of a single bifurcation parameter. On the experimental side, retinoic acid (a morphogene) is an example of a molecule that monotonically increases in concentration when cells traverse the presomitic mesoderm from posterior to anterior. However, the influence of this morphogene on oscillations is currently not (yet) understood on a molecular level, so its relation to our parameter $\alpha$ is therefore open.\\

The second question concerns the very realization of our circuits in the segmentation clock. So far, negative feedback loops were identified in the zebrafish, where certain dimers repress the genes of their components \cite{schroeter2012}, but no indications for self-activating loops have been found so far. This does, of course, not exclude their existence. On the modeling side, for comparison we simulated both, a system of coupled repressilators, and our system of coupled circuits in the limit of very strong coupling. In this case the control of the duration of oscillations failed in the sense that the first and third fixed-point regime either were  absent or the oscillatory regime is replaced by a chaotic one. So our mechanism does not seem to work with simpler building blocks, blocks without positive self-activation.

From a more general perspective  the role of positive feedback loops in combination with negative ones was studied in \cite{tsai}. There it was shown that in general a combination of interlinked positive and negative feedback loops of interacting genes allows to adjust the frequency of a negative feedback oscillator while keeping the amplitude of oscillations nearly constant. At the same time, the combination with a positive feedback loop makes the system more robust against perturbations, and therefore improves its reliability. Although it seems to be open whether these observations are relevant for the segmentation clock, these results together with our observations may trigger a search for self-activating loops on the experimental side of the segmentation clock.

\subsection{Emergence of a self-organized pacemaker}

Among the variety of transient patterns such as spirals, rectangular or plane waves, it is of particular interest that we see also target waves, emitted from units that act like dynamically generated pacemakers, this means in a self-organized way. In an ensemble of oscillators, pacemakers are those units, which entrain the phases of other oscillators to a synchronized oscillation  in space and time. The pacemakers then become the centers of concentric target waves. In our system, the phenomenon of a dynamical generation of a pacemaker is less generic than the generation of spirals, but remarkable as a transient.

\begin{figure}\label{figure17}
	\begin{center}
		\includegraphics[width=8.5cm]{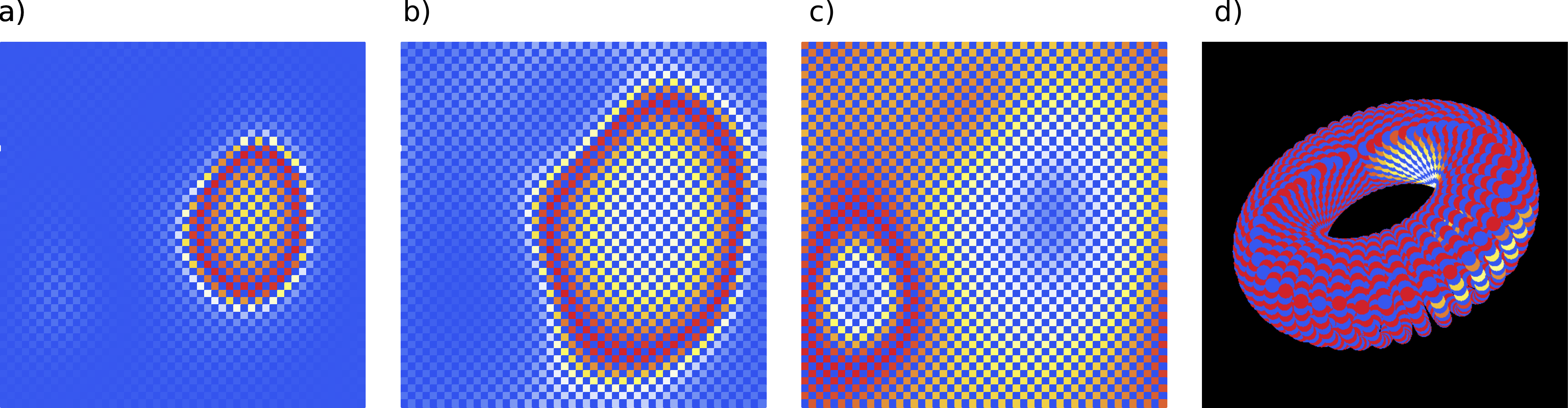}
	\end{center}
	\caption{Snapshots of a pacemaker obtained on an OC-lattice with p.b.c. for slow variation of $\alpha$ and weak coupling at times 14010 (a), 14020 (b), and 14030 (c) and (d), when $\alpha=50$. Panel (a) shows a source of a circular wave that grows (b) and later shrinks into a sink. Both a sink and a source can be seen in (c) and in the corresponding 3d-plot of (d), which shows a snapshot of an evolution of a few periods on a torus. }
\end{figure}

\vskip3pt   Our pacemakers are observed for  the OC-lattice with p.b.c. as transient towards a two-cluster limit cycle either when $\alpha$ is fixed ($\alpha=6$) and $\beta$ is weak, in the first CO-regime of the OC-lattice, or when $\alpha$ is slowly varied  at weak coupling. This choice of the speed within a small interval of allowed variations is essential for the observation.
On the torus topology we then have two centers opposite to each other, one acting as a source of waves, the other one as their sink. The snapshots of Figure~\ref{figure17} display the phase differences of oscillators with respect to the maximal phase value at the time instant of the snapshot. We see roughly circular spots of the same phase differences, located in the bulk of identical phases (blue background in Figure~\ref{figure17}(a)-(b)). The shrinking of the spot corresponds to a sink with respect to the front of constant phase differences, its extension to a source of the same front. In the snapshot of Figure~\ref{figure17}(c)  we see a time instant, for which the two spots, corresponding to the source and the sink with maximal and minimal phase differences are visible at the same time. The corresponding three-dimensional plot is shown in Figure~\ref{figure17} (d). Altogether this collective arrangement of phases lasts over several thousand time units, before this transient approaches a two-cluster CO-state in the form of a chessboard pattern as the stationary state.\\

Rather remarkable about this phenomenon is its emergence in a set of units with a completely uniform  choice of individual parameters ($b,K,\gamma,\alpha$) and coupling $\beta$. Usually pacemakers are implemented with an ad-hoc distinction  via their natural frequencies (see, for example \cite{filippo1}) or some gradient in the natural frequencies of all oscillators \cite{filippo2}; in those cases the location of the pacemaker is predetermined by an ad-hoc implemented distinguished natural frequency. In contrast, our natural frequencies are induced  by a completely uniform choice of parameters in the underlying dynamics. Neither is the final location of the emerging spot distinguished in view of the p.b.c., but the very occurrence of  a pacemaker breaks the symmetry between the oscillatory units. So their heterogeneity must be dynamically generated. Our system, in spite of all couplings being repressive, is of the activator-inhibitor type (as the repression of a repressing bond acts like an activation). Similar self-organized pacemakers have been identified in reaction-diffusion systems \cite{stich1,stich2,stich3}, there, however, as stable phenomena. In those systems the occurrence of self-organized pacemakers could be traced back to the vicinity of a supercritical pitchfork-Hopf bifurcation, in which a difference in frequencies is dynamically generated in this bifurcation. In our case, the conditions for the observation of these transient phenomena and the origin of their instability in terms of bifurcations should be further analyzed in future work.
Although our self-organized pacemakers are transient, their duration over several thousand time units may be sufficient for providing a signal in the biological context, before the system stabilizes to a stationary state.

\section{Summary and Outlook}\label{sec5}

We have studied networks of circuits which individually resemble FitzHugh-Nagumo elements. The three regimes in attractor space of a single unit proliferate to eight regimes if we couple just two of these units. The analysis revealed the interplay of Hopf, pitchfork and fold bifurcations as a function of the bifurcation parameter $\alpha$, the coupling strength $\beta$, and  the system size (where the latter dependence was only indicated). From the numerical integration of the larger systems we found a regime with synchronized oscillations  with spirals and target waves in between regimes with patterns of fixed-point values. Remarkable here was the option for controlling the duration of oscillations via a single parameter. Of much interest for future work is a zoom into the transition region where the bifurcation happens between the different kind of collective behavior. This can be the region where an oscillating ensemble "converts" into an arrested set of units, frozen to their fixed-point values, or vice versa, where oscillations are launched in the "frozen" bulk. We shall analyze the spatio-temporal order of the conversion process in a forthcoming paper.

\subsection*{Acknowledgments}
We thank F. J\"ulicher and D. J\"org (Max-Planck Institute for the Physics of Complex Systems, Dresden) and L.G. Morelli (Universidad de Buenos Aires) for useful discussions about the segmentation clock. Financial support from the Deutsche Forschungsgemeinschaft (DFG, contract ME-1332/19-1) is gratefully acknowledged. We are also indebted to the Galileo Galilei Institute for Theoretical Physics, Arcetri, Florence, where part of this work was done, for its kind hospitality  and partial financial support during our stay at the workshop on "Advances in Non-equilibrium Statistical Mechanics" in May and June 2014.

\appendix

\section{Methods}\label{methods}

Let us start with the nomenclature that we used in the paper. For the option of a {\it Hopf bifurcation} the system has to be at least two-dimensional. When two eigenvalues of a fixed point are complex conjugates and their real parts change sign with the change of a bifurcation parameter, a Hopf bifurcation occurs. A fixed point looses or gains stability and a stable (supercritical) or unstable (subcritical) limit cycle is born, respectively. In particular, in our four-dimensional system it may amount to the lost of stability in all four rather than only in two directions. This can lead to the birth of saddle-limit cycles rather than stable or unstable limit cycles. \\
{\it Fold bifurcations}, here of limit cycles, or synonymously, saddle-node bifurcations of limit cycles, are global bifurcations in which two limit cycles, one stable and one unstable, collide and disappear.\\
A {\it pitchfork bifurcation} involves three fixed points. Depending on the stability of the fixed points, it can be supercritical or subcritical. In supercritical ones, two stable fixed points collide with an unstable fixed point. In the collision, the stable fixed points disappear, and the unstable one becomes stable. In subcritical pitchfork bifurcations, two unstable and one stable fixed point collide, the unstable ones disappear and the stable one becomes unstable.

For a bifurcation analysis our system of four variables is complex due to the relatively large number of variables and the stiffness that is caused by the combination of a fast ($A$) and a slow  ($B$) variable. Therefore we used a combination of the program package  MATCONT, MATHEMATICA  and code written in Fortran 90 in the following way:
\begin{itemize}
\item Firstly we determine all real positive fixed points (since the concentrations of species cannot be negative) for $\alpha \in (0,400)$ and $\beta \in (0,100)$ in steps of $\Delta \alpha = \Delta \beta = 1$ by the program package Wolfram MATHEMATICA, since the convergence by the standard Newton Raphson method  was too slow for small $\alpha$, presumably due to the vicinity of fixed points around the pitchfork bifurcations (the algorithm was then oscillating between different fixed point values).
    \item Next we evaluate the Jacobian from Eqs.~(\ref{eq3}) for $i,j=1,2$ and its eigenvalues at these fixed points using the FORTRAN package LAPACK. The analysis of these eigenvalues reveals the stability of the fixed points and the critical values of $\alpha$, for which the stability of the fixed points changes. These are the $\alpha$-values for which the real part of the eigenvalues changes its sign or they are zero.
    \item Around the critical points we repeat the former steps in a finer resolution of $\Delta \alpha=\Delta\beta=0.01$ or $0.001$ if needed until the real part of the eigenvalues vanishes within the numerical accuracy of $10^{-8}$. To determine the critical points via MATCONT was too slow as compared to the FORTRAN code.
    \item From the critical points we can identify the type of bifurcations, as to whether we have to deal with Hopf or pitchfork bifurcations. Hopf bifurcations are determined as the points, in which the real part of a pair of complex conjugate eigenvalues changes sign. Pitchfork bifurcations happen when three fixed points collide, while two of them disappear  and the third one adopts  the stability of the other two fixed points  from the collision.
    \item In order to decide whether the Hopf bifurcations are subcritical or supercritical, we used MATCONT to calculate the first Lyapunov coefficient.
    \item To confirm the extrapolated behavior in the vicinity of the bifurcation points, we numerically integrated the differential equations  (\ref{eq3}) for different combinations of the bifurcation parameters $\alpha$ and $\beta$. Here we used the second order implicit Runge-Kutta algorithm, since the explicit methods did not converge for all $\alpha$ and $\beta$ values and the variety of initial conditions which we have studied.
    \begin{figure}
	\begin{center}
		\includegraphics[width=8cm]{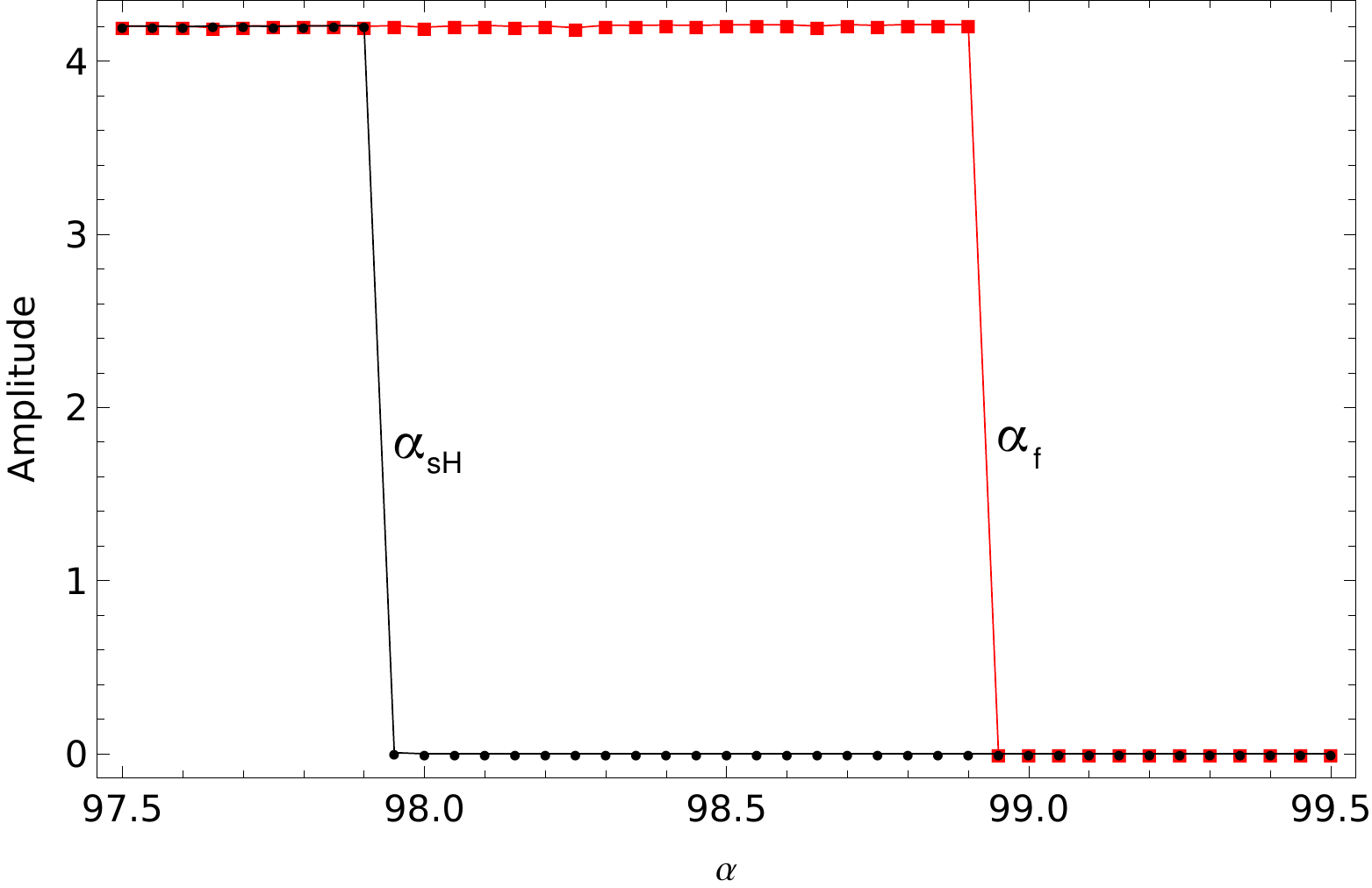}
	\end{center}
	\caption{(Color online) Hysteresis around a subcritical Hopf bifurcation. We change $\alpha$ in small steps starting with $\alpha<\alpha{sH}$ (gray (red) squares), and  with $\alpha>\alpha{f}$ (black dots). The subcritical Hopf bifurcation (gray (red) vertical line) happens at $\alpha_{sH}$, the fold bifurcation of limit cycles (to the accuracy of 0.05) happens at $\alpha_f$. The interval between  the lines represents the coexistence of a limit cycle and a fixed point. The parameters  are $\beta = 0.001$, $\alpha \in (97.5,99.5)$, $\gamma=b=0.01$, $K=0.01$.}\label{figure9}
\end{figure}
    \item To furthermore get access to fold bifurcations of limit cycles which are global bifurcations (that is, not accessible from a linear stability analysis around a fixed point), we numerically integrated the differential equations close to the formerly identified subcritical Hopf bifurcations. We start with $\alpha < \alpha_{sH}$ (the point of the subcritical Hopf bifurcation), where the fixed point becomes unstable and the system evolves to a limit cycle. Once we are sure that the system evolves to a limit cycle, we change the value of $\alpha$ by $0.001$ and use the final state of the integration as an initial condition for the integration with the new value of $\alpha$. We increase $\alpha$ until the system evolves to a stable fixed point. We repeat the procedure from the other side, where $\alpha$ is large enough for the system to evolve to a fixed point from all initial conditions, and decrease $\alpha$ until the system evolves to a limit cycle. Figure~\ref{figure9} shows the amplitude as a function of $\alpha$, that is, either the maximal diameter of the limit cycle or a vanishing amplitude in case of a fixed point. We see the characteristic hysteresis of a subcritical Hopf bifurcation. The first vertical jump occurs at the subcritical Hopf bifurcation at $\alpha_{sH}$, while the second one identifies the fold bifurcation of the limit cycles $\alpha_{f}$, at which the unstable limit cycle collides with the stable limit cycle. In the interval between $\alpha_{sH}$ and $\alpha_{f}$ the stable limit cycle and the stable fixed point coexist. Integrating backwards in time in this interval to obtain numerical values for the amplitude of the unstable limit cycle  was not successful because of the stiffness of the system. We applied this procedure to all subcritical Hopf bifurcations and identified the fold branches $f_1$ and $f_2$ in this way.
We did not find any other bifurcations for these intervals of $\alpha$ and $\beta$.
\end{itemize}

\section{Bifurcation analysis along lines of constant $\beta$ of Figure 2 from the main text}\label{det_bif}
\noindent This presentation is more detailed in the sense that we refer to concrete numbers as compared to Sec.~\ref{secbifurcation} of the main text.

\subsection{The $\beta = 0.001$-section} \label{sec:beta0.001_line}
The $\beta=0.001$-line corresponds to the $L_1$-line in Figure~\ref{figure7} and describes a scenario that is typical for $\beta \in (0,0.003)$. For this choice of parameters the system has only one real positive fixed point. The fixed point first looses and later gains its stability through subcritical Hopf bifurcations. There are in total four subcritical bifurcations, two around $\alpha = 30$ and two around $\alpha = 98$, they are either double subcritical Hopf bifurcations or very close to each other. Figure~\ref{figure8} shows a sketched diagram of how the limit cycles are created and destroyed. For $\beta = 0.001$ and $\alpha = 30.1$ the first double subcritical Hopf bifurcation $h_1$ occurs, this creates an unstable limit cycle (dashed black curve in Figure~\ref{figure8}(a)) that collides with the large stable limit cycle (full black line) at the fold bifurcation of limit cycles $f_2$. This stable limit cycle exists between the two fold bifurcations $f_1$ and $f_2$. At $\alpha = 98$ the second double subcritical Hopf bifurcation $h_2$ happens, when a new unstable limit cycle is born. This limit cycle also collides in a fold bifurcation $f_1$ with the large stable limit cycle, see Figure~\ref{figure8}.

\subsection{The $\beta = 0.01$-section} \label{sec:beta0.01_line}
For $\beta > 0.003$ the double Hopf bifurcation is sufficiently split so that we can numerically distinguish two separate Hopf bifurcations $h_{11}$ and $h_{12}$. The second Hopf bifurcation $h_{12}$ becomes supercritical. This means that the limit cycle that is born in this way is stable, in this case it is a saddle-limit cycle. Figure~\ref{figure8}(b) shows a bifurcation diagram for varying $\alpha$ at $\beta=0.01$, this corresponds to an $L_2$-line in Figure~\ref{figure7}. The homogeneous fixed point looses its stability in two directions through a subcritical Hopf bifurcation $h_{11}$. An unstable limit cycle is created, the attractor of the system is a large relaxation cycle, which is created and destroyed in the same way as for $\beta=0.001$. Soon after this bifurcation, a supercritical Hopf bifurcation $h_{12}$ occurs, when a saddle-limit cycle is created (black dotted line in Figure~\ref{figure8}(b) and (c)) and the fixed point  looses its stability in all directions. This limit cycle is attractive from two directions and repelling from the other two directions. The relatively high dimensionality of the system makes the calculations tedious; since we are only interested in a qualitative description, we did not attempt to pursue the attractive and repulsive directions any further. At $\alpha = 98$ a double subcritical Hopf bifurcation $h_2$ occurs as described in section~\ref{sec:beta0.001_line}, an unstable limit cycle is created, which collides with both the stable and the saddle-limit cycle via a fold bifurcation of limit cycles, see Figures \ref{figure8}(b) and (c).

\subsection{The $\beta = 0.1$ section} \label{sec:beta0.1_line}
At the point ($\alpha,\beta$) = (0,0.04) (point $h_3$ in Fig.~\ref{figure8}(d)) three additional bifurcation branches emerge. So for $\beta = 0.1$ and small $\alpha$ the system has three positive fixed points, two symmetric stable, and one homogeneous unstable. Two stable fixed points have negative real parts of all eigenvalues, while the unstable fixed point has three negative and one positive real part of the four eigenvalues. The system evolves to one of the stable fixed points, depending on the initial conditions. The symmetric fixed points loose their stability at the same time, at $\alpha= 1.69$, when the real parts of two complex conjugate eigenvalues change their sign and become positive. These are the subcritical Hopf bifurcations $h_{31}$ and $h_{32}$ (for which the first Lyapunov coefficient is positive). An unstable limit cycle is born, it collides with the stable limit cycle, which was born through the supercritical Hopf bifurcation $h_{33}$ that happens at $\alpha = 4.0088$, see Figure~\ref{figure8}(e).

The collision of the limit cycles through a fold bifurcation of limit cycles $f_3$ happens at $\alpha = 1.67$. For the interval of $\alpha$ between $1.67$ and $1.69$ we find a coexistence of three attractors, two fixed points and one limit cycle. In the interval between $1.69$ and $4.0088$ the only attractor is a limit cycle. In this interval at $\alpha = 2.13$ we have a pitchfork bifurcation, at which all three fixed points collide, see Figure~\ref{figure8}(f). Before the collision the two symmetric fixed points had two positive and two negative eigenvalues, and the third fixed point had three negative and one positive eigenvalue. The two symmetric fixed points disappear in the collision and the third one changes is stability, now it has two positive and two negative eigenvalues (like the two stable fixed points before the collision). This bifurcation does not affect the attractor (a limit cycle) in this interval, since none of the fixed points involved in the bifurcation is stable, before and after the bifurcation.

In the interval between $4.0088$ and $28.9$ the surviving fixed point from the pitchfork bifurcation is stable (all eigenvalues are negative), and it is the only attractor in this interval. At $\alpha = 28.9$ there is a subcritical Hopf bifurcation, with the first Lyapunov coefficient being positive, the fixed point looses its stability when the real parts of two eigenvalues change their sign. An unstable limit cycle is born, the attractor of the system is a large stable limit cycle. At $\alpha = 32.4$ the real parts of the other two eigenvalues change sign, and the fixed point now has all eigenvalues negative. This is a supercritical Hopf bifurcation, a saddle-limit cycle is created.

In the interval between $32.4$ and $97.92$ the two limit cycles coexist, depending on the initial conditions, the system will immediately evolve to a large stable limit cycle or first to a saddle-limit cycle, where it spends a long time before it evolves to the large stable limit cycle. Both limit cycles disappear through a fold bifurcation of limit cycles at $\alpha = 99$, when they collide with the unstable limit cycle that is created by the subcritical Hopf bifurcation $h_2$ at $\alpha = 97.94$. Between $\alpha= 97.94$ and $99$ we have a coexistence of the two limit cycles and a fixed point. After the fold bifurcation the fixed point remains as the only attractor.

\subsection{The $\beta = 10$-section} \label{sec:beta10_line}

For $\beta = 10$ we have no longer the $h_1$-branch from Figure 2 in the main text. For smaller $\alpha \in (0,22.63)$ we have three fixed points, two stable and one unstable (with three stable and one unstable directions), se Figure~\ref{figure8}(h). The two stable fixed points loose their stability in two directions through a subcritical Hopf bifurcation. In the interval $\alpha \in (20.7,22.63)$ we have a coexistence of the two stable fixed points and one stable limit cycle. The unstable limit cycle created in the subcritical Hopf bifurcation collides with this large stable limit cycle in a fold bifurcation of limit cycles at $\alpha = 20.7$. Between the first subcritical Hopf bifurcation and the supercritical Hopf bifurcation at $\alpha = 42.69$ the only attractor is a stable limit cycle. At the supercritical bifurcation a saddle limit cycle is born. Here as in all cases before, the larger stable limit cycle and the smaller saddle-limit cycle differ not only in their amplitudes, which for small $\beta$ can be very similar, but also in their frequency and  synchronization pattern. The large stable limit cycle is a two-cluster limit cycle, while the small saddle-limit cycle is a one-cluster limit cycle, cf. Figure 4 in the main text. The saddle-limit cycle created through the supercritical Hopf bifurcation experiences a canard explosion, for a very small increase of $\alpha$ it jumps from a small limit cycle to a large relaxation cycle, see Figure 5 in the main text. This phenomenon is very common for systems with fast and slow variables. At $\alpha = 96.14$ an unstable limit cycle is born that almost immediately collides with the saddle-limit cycle, see Figure~\ref{figure8}(i).

Figure 3 in the main text shows a bifurcation diagram of eigenvalues (Figure 3(a) in the main text) and of fixed points and limit cycles (Figure 3(b) in the main text) for $\beta=10$ and varying $\alpha$. One can compare the sketched diagram in Figure~\ref{figure8}(h) with real data in Figure 3(b) in the main text. At $\alpha = 97.94$ we see the last subcritical Hopf bifurcation $h_{22}$, when an unstable limit cycle is born. This limit cycle collides  with the large stable limit cycle at $\alpha = 251$ in a fold bifurcation of limit cycles $f_1$, see Figure~\ref{figure8}(i) and 3(b) in the main text. Afterwards only a homogeneous stable fixed point remains as attractor.

\subsection{The $\beta = 100$-section} \label{sec:beta100_line}

For $\beta > 76.84$ the $h_2$-branch of Figure 2 in the main text does no longer exist, so the smaller one-cluster limit cycle ceases to exist for any $\alpha$. As before, for small values of $\alpha$ there are two stable and one unstable fixed points. Stable fixed points remain attractors until they loose their stability through a subcritical Hopf bifurcation $h_{3}$. The unstable limit cycle, born at this point, collides with the two-cluster stable limit cycle in a fold bifurcation of limit cycles, when they both disappear. After the pitchfork bifurcation, the surviving  fixed point has two positive and two negative real parts of eigenvalues. The two positive change sign at the subcritical Hopf bifurcation $h_{22}$, when the fixed point becomes stable, and an unstable limit cycle is born. This unstable limit cycle collides with the stable one at a second fold bifurcation of limit cycles, see Figure~\ref{figure8}(j). In the interval before the first fold bifurcation the attractors are two fixed points, in the interval between the first fold and the first Hopf bifurcations, the coexisting attractors are two fixed points and one limit cycle, between the first and the second Hopf bifurcations the attractor is a limit cycle, between the second Hopf and the second fold bifurcation the attractors are a fixed point and a limit cycle, and after the second fold bifurcation the attractor is a fixed point.

\section{Extrapolation of bifurcations towards larger system sizes}\label{4units}
\begin{figure}[ht]
\center
\includegraphics[width=5 cm]{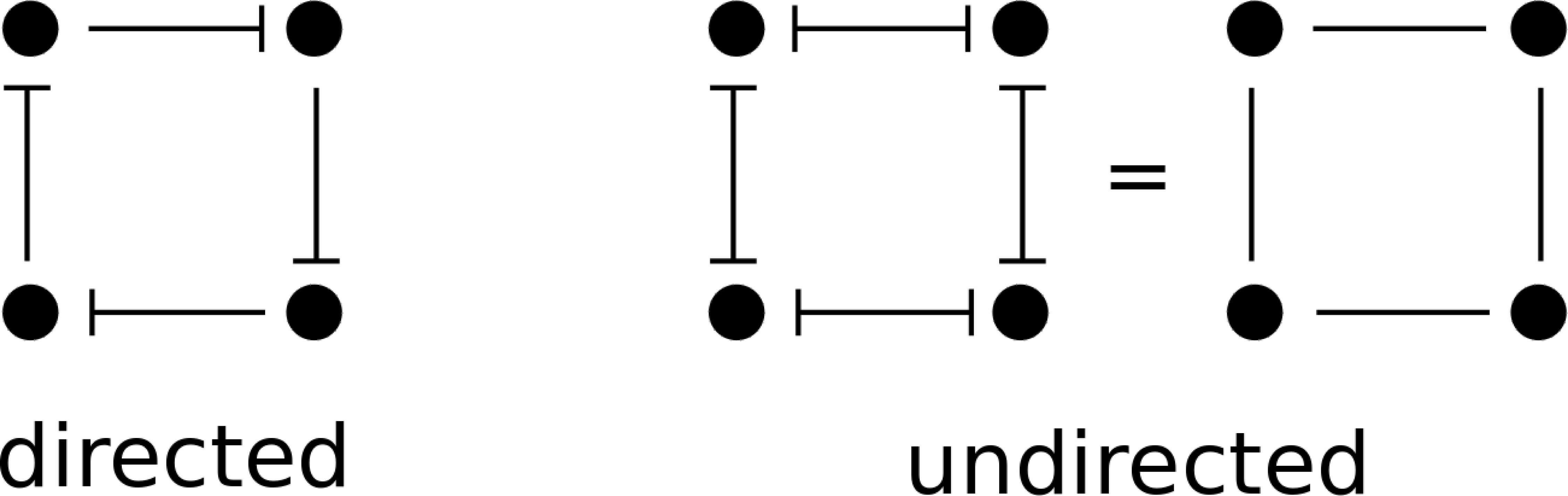}
\caption{Four coupled circuits, directed (left) and undirected (right).}\label{figure12}
\end{figure}

For a single genetic circuit we  found three regimes with qualitatively different behavior: fixed-point behavior for small and large $\alpha$, differing by the fixed-point values, and limit-cycle behavior for an intermediate range of $\alpha$. For two coupled circuits we found already eight regimes, as described in the previous section. In a naive extrapolation one may expect a proliferation of regimes in larger sets of coupled circuits. On the other hand, for the $50\times 50$ sets of coupled units we observe a relatively simple structure of mainly three regimes as it is outlined in the section IV of the main text. Therefore it is instructive to illustrate with four units, coupled with directed or undirected links according to Figure~\ref{figure12}, how  multiple Hopf bifurcations may be generated for a larger number of units on the one hand, while on the other hand distinct bifurcation branches may merge for a different choice of couplings. So it is inherently difficult, if not impossible, to extrapolate the collective behavior to larger system sizes.

We couple four units in two different ways: either with directed coupling, where one unit represses one of the neighbors and gets repressed by the other one, or with undirected couplings, where all neighbors repress each other mutually in the same way and with the same strength, see Figure~\ref{figure12}. Using MATCONT we check bifurcations for the homogeneous fixed point that survives to large system sizes (with all variables having the same value) for values of $\beta = 0.01, 0.1, 1.0, 10.0$, and $100.0$ and varying $\alpha$.

\begin{figure}[ht]
\center
\includegraphics[width=6 cm]{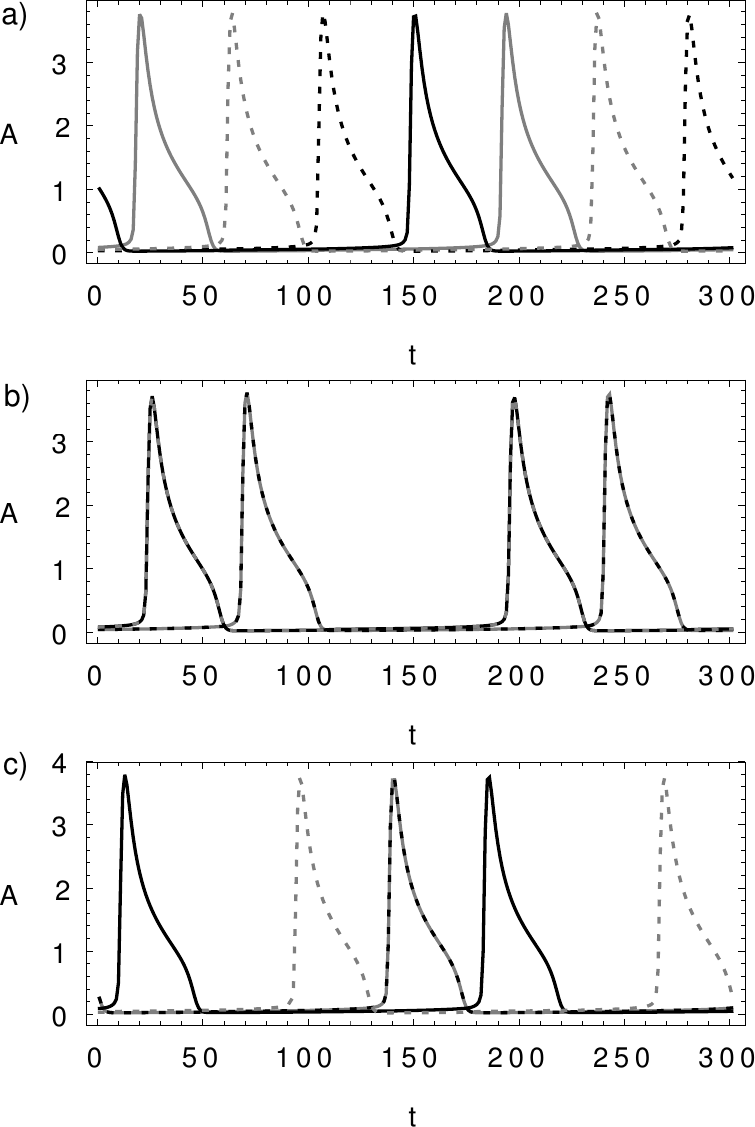}
\caption{Three different limit cycles for four synchronized undirected units, differing by their pattern of phase-locked motion. The parameter values are $\beta = 0.01$, $\alpha = 60$, $\gamma=b=0.01$, $K=0.01$.}\label{figure13}
\end{figure}

For both topologies, we find a splitting of the two Hopf bifurcations at small and large $\alpha$ into three or four Hopf bifurcation branches for $\beta > 0$. These branches surround regimes of different attractors. Some of these regimes have more than two coexisting attractors, see Figure~\ref{figure13}. In both cases, as in the case of the toggle switch, we find for $\beta \geqslant 0.1$ coexisting fixed points, which disappear in a pitchfork bifurcation, after which only one fixed point is left. The stability of the surviving fixed point depends on the value of $\alpha$. In the regime, where the fixed point is unstable, the system evolves to one of the limit cycles. In both cases (directed and undirected), the Hopf bifurcation at the larger value of $\alpha$ splits into three branches. One of the three branches, the middle one, is a double Hopf bifurcation in the sense that the real part of two pairs of complex conjugate eigenvalues changes sign at the same time. For sufficiently large $\beta$, only one of the three branches survives. The Hopf bifurcation at the smaller values of $\alpha$ splits into four (directed) or three (undirected) branches. All lower branches disappear for $\beta$ sufficiently large, for the undirected case this happens for $\beta < 100$ and for directed couplings for $\beta > 100$. After all Hopf branches apart from the one at the largest value of $\alpha$ disappear, the system has only one limit cycle in an oscillatory regime, a two-cluster limit cycle, which we also see as a chessboard pattern on a larger lattice, see Figure 15(c) in the main text.\\
From the Jacobian of a system of $N$ units, evaluated at the homogeneous fixed point,  one can read off that the number of Hopf bifurcations increases sub-linearly in $N$. For $N$ units we can have up to $N$-fold Hopf bifurcations with N pairs of complex conjugate eigenvalues whose real part changes sign at the same time (as we saw for the two and four units). These bifurcations start to separate as we increase $\beta$. Some of these branches may be sufficiently separated  to create numerically distinguishable  regimes with multistability, but for a different choice of couplings they can merge and reduce the number of different regimes. The case of four directed couplings had been studied in
\cite{pablo} where it was shown that a plaquette of four circuits (there termed bistable frustrated units) is multistable for a certain choice of parameters in agreement with our results here (see Figure~\ref{figure13}).

\end{document}